\documentclass[reqno,11pt]{amsart}
\usepackage{graphicx}
\usepackage{amscd}
\usepackage{slashed}
\usepackage{amssymb}
\usepackage{esint}
\usepackage[off]{auto-pst-pdf} 
\usepackage{pst-grad} 
\usepackage{pst-plot} 
\usepackage[mathscr]{eucal}
\numberwithin{equation}{section}
\allowdisplaybreaks[1]

\newcommand{\SetFigFont}[3]{}

\title[The Fermionic Signature Operator in an Electromagnetic Wave]{The Fermionic Signature Operator and \\
Hadamard States in the Presence of a \\ Plane Electromagnetic Wave}

\author[F.\ Finster]{Felix Finster}
\address{Fakult\"at f\"ur Mathematik \\ Universit\"at Regensburg \\ D-93040 Regensburg \\ Germany}
\email{finster@ur.de}

\author[M.\ Reintjes]{Moritz Reintjes \\ \\ September 2016 / January 2017}
\address{Departamento de Matem{\'a}tica \\ Instituto Superior T{\'e}cnico \\
1049-001 Lisboa \\ Portugal}
\email{moritzreintjes@gmail.com}

\newtheorem{Def}{Definition}[section]
\newtheorem{Thm}[Def]{Theorem}

\newtheorem{Lemma}[Def]{Lemma}

\newtheorem{Corollary}[Def]{Corollary}

\newcommand{\Thanks}{\vspace*{.5em} \noindent \thanks}
\newcommand{\beq}{\begin{equation}}
\newcommand{\eeq}{\end{equation}}
\newcommand{\Proof}{\begin{proof}}
\newcommand{\QED}{\end{proof} \noindent}

\newcommand{\la}{\langle}
\newcommand{\ra}{\rangle}
\newcommand{\bra}{\mathopen{<}}
\newcommand{\ket}{\mathclose{>}}
\newcommand{\Sl}{\mathopen{\prec}}
\newcommand{\Sr}{\mathclose{\succ}}

\newcommand{\C}{\mathbb{C}}
\newcommand{\R}{\mathbb{R}}
\newcommand{\1}{\mbox{\rm 1 \hspace{-1.05 em} 1}}
\newcommand{\Z}{\mathbb{Z}}
\newcommand{\N}{\mathbb{N}}

\renewcommand{\H}{\mathscr{H}}

\newcommand{\bep}{\begin{pmatrix}}
\newcommand{\enp}{\end{pmatrix}}

\newcommand{\Dir}{{\mathcal{D}}}
\newcommand{\D}{{\mathscr{D}}}

\newcommand{\B}{{\mathscr{B}}}

\newcommand{\Lin}{\text{\rm{L}}}

\newcommand{\Cisc}{C^\infty_{\text{sc}}}
\newcommand{\Cisco}{C^\infty_{\text{\rm{sc}},0}}
\newcommand{\n}{\mathfrak{n}}

\DeclareMathOperator{\supp}{supp}

\newcommand{\p}{\mathfrak{p}}
\newcommand{\Sig}{\mathscr{S}}
\newcommand{\scrM}{\mycal M}
\newcommand{\scrN}{\mycal N}

\DeclareFontFamily{OT1}{rsfso}{}
\DeclareFontShape{OT1}{rsfso}{m}{n}{ <-7> rsfso5 <7-10> rsfso7 <10-> rsfso10}{}
\DeclareMathAlphabet{\mycal}{OT1}{rsfso}{m}{n}

\begin{document}

\begin{abstract}
We give a non-perturbative construction of a distinguished state for
the quantized Dirac field in Minkowski space in the presence of a time-dependent external field
of the form of a plane electromagnetic wave.
By explicit computation of the fermionic signature operator, it is 
shown that the Dirac operator has the strong mass oscillation property.
We prove that the resulting fermionic projector state is a Hadamard state.
\end{abstract}

\maketitle

\tableofcontents

\section{Introduction}
In the canonical formulation of quantum field theory (QFT), the physical system is described by
a vector of the Fock space, which in turn is built up from a ground
state\footnote{Here by ``ground state'' we mean any state
taken as the starting point of the Fock space construction.}
by applying the creation
operators of the particle and anti-particle states. For Dirac fermions in Minkowski space,
a distinguished ground state is given by the vacuum state, whose construction involves a {\em{frequency splitting}}.
In simple terms, one decomposes the solution space of the Dirac equation into
the subspaces of solutions of positive and negative frequency (these subspaces can be
described equivalently as the positive and negative spectral subspaces of the Dirac Hamiltonian).
After reinterpreting the annihilation operators of the negative-frequency solutions as creation operators
and vice versa, the vacuum state
is defined as the unit vector which vanishes when acted upon by
the annihilation operators (for a formulation of this construction in the language of algebraic QFT
see Section~\ref{secquant} below).

In this paper, we are concerned with the basic problem that in the {\em{time-dependent setting}},
there is no obvious choice for the ground state because frequency splitting can no longer be used.
More specifically, we consider a massive Dirac field in Minkowski space
in the presence of an {\em{external electromagnetic field}} of the form of a {\em{plane wave}}
(i.e.\ an electromagnetic potential depending only on~$t+x$).
The physical situation we have in mind is that of a laser beam whose amplitude and frequency
may be arbitrarily large.
If one considers a short laser pulse, then the physical scenario can be described as a scattering process,
with a unique ground state (= the vacuum state) before and after the interaction with the laser pulse
takes place. However, in this setting of a scattering process,
it is unclear how the quantum state is to be described while the interaction is present.
The reason is that our system is strongly interacting, so that notions like ``perturbations of
states on the mass shell'' are questionable.
The situation becomes even more involved if, instead of a short laser pulse, one considers a laser
beam present for all times, possibly with the amplitude and frequency changing in time.
It is the main objective of this paper to show that even in such time-dependent situations
without a prescribed asymptotics for large times, there is a distinguished ground state
which is of Hadamard form.

We now give some more background on the problem and mention related works.
The basic problem that frequency splitting cannot be used to obtain a decomposition
of the Dirac solution space into two subspaces is sometimes referred to as the external field problem
(see for example~\cite[\S2.1.2]{cfs}). As a result, in the second-quantized description with {\em{Fock spaces}},
it is not only unclear which ground state to choose, but even more the ground state
obtained from the negative spectral subspace of the Hamiltonian at a fixed time
is no longer a vector of the Fock space built up in the Minkowsi vacuum or built up from
the ground state at another time.
This was first observed by Klaus and Scharf~\cite{klaus+scharf1, klaus+scharf2} in the presence of a
static external potential, and was later analyzed by Fierz in Scharf~\cite{fierz+scharf} in the time-dependent setting.
This fact has severe consequences for the time evolution in Fock spaces. The mathematical results of
Shale-Stinespring~\cite{shale+stinespring} and Ruijsenaars~\cite{ruijsenaars}
show that a second-quantized time evolution cannot be
implemented on a fixed Fock space, unless the spatial components of the electromagnetic potential are zero.
A recent approach to tackle this problem is to work with time-dependent Fock spaces~\cite{merkl, deckert+merkl2}.
Generally speaking, the freedom in choosing the ground state for the Fock space construction
implies that the interpretation of the physical state in terms of particles and anti-particles depends
on the observer. Thus, describing experiments involving particles or anti-particles,
one must specify a physical model of the detector.

An alternative approach, which has proved to be fruitful in
algebraic QFT in curved space-time, is provided by {\em{microlocal analysis}}.
Here one makes use of the fact that for a free Dirac field,
the quantum state is uniquely characterized by its two-point function (using Wick's theorem),
being a bi-distribution in space-time.
Characterizing the singularity structure of this two-point distribution with notions of
microlocal analysis gives rise to the {\em{Hadamard form}}
(for details see Section~\ref{sechadamard}).
The expansion of the two-point distribution in the order of the
singularity on the light cone is called {\em{Hadamard expansion}}.
In the setting of Minkowski space to be considered here, the
Hadamard expansion agrees with the {\em{light-cone expansion}}
(see~\cite{light} or the introduction in the textbook~\cite[Section~2.2]{cfs}).
A quantum state whose two-point distribution is of Hadamard form is
called {\em{Hadamard state}}.
For Hadamard states, correlations of powers of Wick-ordered products
are well-defined, making it possible to build up a perturbative QFT (see for
example~\cite{fewster2013necessity, dappiaggiDirac}
or the recent text book~\cite{rejzner}). This is why Hadamard states are generally
accepted as a class of physically sensible quantum states.
Just as in the Fock space formalism, in the microlocal approach the particle interpretation depends
on the observer. This is reflected mathematically by the fact that the Hadamard condition
specifies the two-point distribution only up to smooth contributions.
For a good introduction to free quantum fields in curved space-time
we refer to~\cite{waldQFT, hollands+wald} or~\cite{baer+fredenhagen}.

The {\em{fermionic signature operator}} introduced in~\cite{finite, infinite}
(based on earlier perturbative constructions in~\cite{sea}) provides a distinguished
fermionic ground state in space-time, referred to as the fermionic projector (FP) state.
Here by ``distinguished'' we mean that the construction is covariant
and gives rise to a unique state. In particular, the construction does not involve
the choice of coordinates, gauges or reference frames.
In an ultrastatic space-time of infinite lifetime, the FP state coincides with the
ground state obtained from the usual frequency splitting in the reference frame
of an observer at rest (see~\cite[Section~5]{sea} and~\cite[Section~5]{infinite}).
In the time-dependent setting, however, the
FP state cannot be associated to a local observer. Instead, it depends on the global
geometry of space-time (for a more detailed account on the physical interpretation
we refer to~\cite[Section~2.1.2]{cfs} or the discussion of a scattering process in~\cite[Section~5]{sea}).
In mathematical terms, the fermionic signature operator is a symmetric operator
on the solution space of the massive Dirac equation
in globally hyperbolic space-times. It encodes geometric information~\cite{drum} and 
gives a new covariant method for obtaining Ha\-da\-mard states~\cite{hadamard}.
The abstract construction in space-times of
finite and infinite lifetime as given in~\cite{finite, infinite} opens up the research program
to explore the fermionic signature operator in various space-times and to verify if the resulting
FP states are Hadamard.
So far, the fermionic signature operator has been studied in the examples 
of closed FRW space-times~\cite{finite},
ultrastatic space-times of infinite lifetime and de Sitter space-time~\cite{infinite},
ultrastatic slab space-times~\cite{fewster+lang}
as well as for an external potential in Minkowski space which decays at infinity~\cite{hadamard}.
Moreover, as the first example involving a horizon, the Rindler space-time is considered in~\cite{rindler}.
Finally, in~\cite{drum} various two-dimensional examples are analyzed.
In the example of a space-time slab, it is shown in~\cite{fewster+lang} that the resulting FP state
is in general not of Hadamard form, but that one gets Hadamard states if the construction is ``softened''
by a smooth cutoff function in time. On the other hand, in ultrastatic space-times of infinite
lifetime~\cite[Section~5]{infinite}, the two-dimensional Rindler space-time~\cite[Section~11]{rindler} as well as for
an external potential in Minkowski space~\cite[Theorem~1.3]{hadamard}, the resulting FP state
is proven to be of Hadamard form. These results lead to the conjecture that the FP state should be of
Hadamard form provided that space-time is ``sufficiently smooth'' on its boundaries and at its asymptotic ends.
In order to challenge and quantify this conjecture, one needs to construct and analyze the fermionic signature
operator in detail in different space-times.

In the present paper, the fermionic signature operator is constructed for the first time in
the presence of an external potential which is neither static nor decays for large times.
After the necessary preliminaries (Section~\ref{secprelim}),
the Dirac equation in the presence of a plane electromagnetic wave is considered (Section~\ref{secEMwave}).
Clearly, using the separation ansatz
\beq \label{sepintro}
\psi (t,x,y,z) = e^{-i k_2 y - i k_3 z}\: e^{-iu(t-x)} \: \chi_{k_2, k_3,u}(t+x) \:,
\eeq
the Dirac equation can be reduced to a system of ODEs. Moreover, since~$t+x$ is a lightlike direction,
the resulting scalar ODE is of first order, making it possible to solve the equation explicitly by integration.
This structure suggests that one should consider the Dirac equation as an evolution
equation which maps initial data on the null surface~$t+x=\text{const}_1$ to another such
surface~$t+x=\text{const}_2$. In order to establish such a formulation, one needs to show that the
spatially compact solutions of the Dirac equation have suitable decay properties when restricted to the
surfaces~$t+x=\text{const}$. This decay in null directions is derived in Section~~\ref{secnull}.
The first step is to show that for fixed~$k_2$ and~$k_3$, the Dirac equation in the~$tx$-plane
can be reduced to a two-dimensional Klein-Gordon equation (without external potentials) by transforming
to a suitable curvilinear coordinate system (see~\eqref{KG2}). This makes
it possible to apply well-known decay results for the massive Klein-Gordon equation in Minkowski space.
We thus obtain rapid decay in null directions (Lemma~\ref{lemmanull}), and we can rewrite
the scalar product on the solution space in terms of integrals over the null surfaces~$t+x=\text{const}$
(Lemma~\ref{spatial_int_Lemma}). In Section~\ref{secsig}, it is shown that the Dirac equation has the strong
mass oscillation property and that the resulting fermionic signature operator simply is the multiplication operator by the
sign of the separation constant~$u$ in~\eqref{sepintro} (Theorem~\ref{thmsig}).
In Section~\ref{secquant}, we review the construction of the fermionic projector
and the FP state (Theorem~\ref{thmstate}).
The just-mentioned result of Theorem~\ref{thmsig}
can be understood immediately from the fact that without electromagnetic
potential, the variable~$u$ is positive on the upper mass shell and negative on the lower mass shell,
giving agreement with the fermionic signature operator as computed in~\cite[Section~3]{hadamard}.
Moreover, the FP state coincides with the state
obtained by frequency splitting in the momentum variable~$u$ in~\cite[Section~5.2]{fradkin}.
One should keep in mind that, if an electromagnetic potential is present, solutions for negative~$u$ may well have
arbitrarily large positive frequencies (as is illustrated in Figure~\ref{figharmonic}).
The delicate point of our analysis is to show that, despite this effect, the FP state is a Hadamard state
(Corollary~\ref{corH}). For the proof, we first derive an integral representation of the fermionic
projector (Section~\ref{secfourier}) and study it with methods of microlocal analysis
(Section~\ref{sechadamard}).
Our constructions and results are illustrated
in Section~\ref{secexharmonic} in the explicit example of a harmonic plane electromagnetic wave~\eqref{Aharmonic}.
It is a main advantage of our construction that it also applies in situations without symmetries
as obtained for example by adding an electromagnetic potential with suitable decay properties
at infinity. This is explained in Section~\ref{secoutlook}.

\section{Preliminaries} \label{secprelim}
Let~$\scrM$ be Minkowski space, a four-dimensional real vector space endowed with an
inner product of signature~$(+ \ \!\! - \ \!\! - \ \! - )$.
We let~$S\scrM$ be the spinor bundle on~$\scrM$ and denote the smooth sections of the spinor bundle by~$C^\infty(\scrM, S\scrM)$. Similarly, $C^\infty_0(\scrM, S\scrM)$ denotes the smooth sections with compact support. The fibres~$S_p\scrM$ are endowed with an inner product of signature~$(2,2)$ which we denote by~$\Sl .|. \Sr_p$. An external potential is a multiplication operator~$\B(p) \in \Lin(S_p\scrM)$, which we assume to
be smooth and symmetric with respect to the spin scalar product, i.e.
\[ \B \in C^\infty(\scrM, \Lin(S\scrM)) \qquad \text{with} \qquad
\Sl \B \phi | \psi \Sr_p = \Sl \phi | \B \psi \Sr_p \quad \forall \phi, \psi \in S_p\scrM\:. \]
The Dirac matrices~$\gamma^j$, $j=0,...,3$, are symmetric linear operators on~$(S_p\scrM, \Sl .|. \Sr_p)$
which satisfy the anti-commutation relations
\beq \label{anti-com}
\{ \gamma^i,\gamma^j\} \equiv \gamma^i\gamma^j + \gamma^j\gamma^i = 2\,\eta^{ij} \,\1_{S_p \scrM}\:,
\eeq
where~$\eta = \text{diag}(1,-1,-1,-1)$ is the Minkowski metric.
The Dirac operator is defined by
\beq \label{Dirdef}
\Dir := i \gamma^j \partial_j + \B \::\: C^\infty(\scrM, S\scrM) \rightarrow C^\infty(\scrM, S\scrM)\:.
\eeq
For a given real parameter~$m>0$ (the ``rest mass''), the Dirac equation reads
\beq \label{dirac}
(\Dir - m) \,\psi_m = 0 \:.
\eeq

In the Cauchy problem, one seeks for a solution of the Dirac equation with
initial data~$\psi_\scrN$ prescribed on a given Cauchy surface~$\scrN$. Thus in the smooth setting,
\[ (\D - m) \,\psi_m = 0 \:,\qquad \psi_m|_{\scrN} = \psi_\scrN \in C^\infty(\scrN, S\scrM) \:. \]
This Cauchy problem has a unique solution~$\psi_m \in C^\infty(\scrM, S\scrM)$.
This can be seen by considering energy estimates for symmetric hyperbolic systems
or by constructing the Green's kernel (see for example~\cite{john} and~\cite{baer+ginoux}). These methods
also show that the Dirac equation is causal,
meaning that the solution of the Cauchy problem only depends on the initial data in the causal
past or future. In particular, if~$\psi_\scrN$ has compact support, the solution~$\psi_m$ also has compact
support on any other Cauchy hypersurface. This leads us to consider solutions~$\psi_m$
in the class~$\Cisc(\scrM, S\scrM)$ of smooth sections with spatially compact support. On solutions in this class,
one introduces the scalar product~$(.|.)_\scrN$ by
\beq \label{print}
(\psi_m | \phi_m)_\scrN = 2 \pi \int_\scrN \Sl \psi_m | \gamma^j \nu_j\, \phi_m \Sr_p\: d\mu_\scrN(p) \:,
\eeq
where~$\nu$ is the future-directed normal.
Due to current conservation, this scalar product does not depend on the choice of the Cauchy surface~$\scrN$
(for details see~\cite[Section~2]{finite}).
Therefore, we may omit the subscript~$\scrN$ and simply denote
the scalar product~\eqref{print} by~$( .|. )$. Forming the completion, we obtain
the Hilbert space~$(\H_m, (.|.))$.

The {\em{retarded}} and {\em{advanced Green's operators}}~$s_m^\wedge$ and~$s_m^\vee$ are
mappings (for details see for example~\cite{baer+ginoux})
\[ s_m^\wedge, s_m^\vee \::\: C^\infty_0(\scrM, S\scrM) \rightarrow \Cisc(\scrM, S\scrM)\:. \]
Their difference is the so-called causal fundamental solution~$k_m$,
\beq \label{kmdef}
k_m := \frac{1}{2 \pi i} \left( s_m^\vee - s_m^\wedge \right) \::\: C^\infty_0(\scrM, S\scrM) \rightarrow \Cisc(\scrM, S\scrM) \cap \H_m \:.
\eeq
This operator can be represented as integral operators with a distributional kernel,
\[ (k_m \phi)(x) = \int_\scrM k_m(x,y)\, \phi(y)\: d^4y\:. \]

For the following construction, the mass parameter in the Dirac equation~$m$ will not be fixed.
Instead, it can vary in an open interval~$I:=(m_L, m_R)$ with~$m_L, m_R>0$.
We denote the families of smooth wave functions with spatially compact support, which are also compactly supported in~$I$, by~$\Cisco(\scrM \times I, S\scrM)$.
The space of families of Dirac solutions in the class~$\Cisco(\scrM \times I, S\scrM)$
are denoted by~$\H^\infty$. On~$\H^\infty$ we introduce the scalar product
\[ ( \psi | \phi) = \int_I (\psi_m | \phi_m)_m \: dm\:, \]
where~$dm$ is the Lebesgue measure (and~$\psi = (\psi_m)_{m \in I}$
and~$\phi = (\phi_m)_{m \in I}$ are families of Dirac solutions for a variable mass parameter).
Forming the completion yields the Hilbert space $(\H, (.|.))$ with norm~$\| . \|$.
Then~$\H^\infty$ can be regarded as the subspace
\beq \label{Hinfchoice}
\H^\infty = \Cisco(\scrM \times I, S\scrM) \cap \H \:.
\eeq
Integrating~$\psi_m$ over~$m$ gives the operator
\[ \p \::\: \H^\infty \rightarrow \Cisc(\scrM, S\scrM)\:,\qquad
\p \psi = \int_I \psi_m\: dm \:. \]
For clarity, we point out that~$\p \psi$ no longer satisfies a Dirac equation.
Finally, on Dirac wave functions (not solutions of the Dirac equation)
we introduce the Lorentz invariant inner product 
\beq \label{stipMin}
\bra \psi|\phi \ket = \int_\scrM \overline{\psi(x)} \phi(x) \: d^4x \:,
\eeq
whenever the integral on the right converges.

The following notion was introduced in~\cite{infinite}, and we refer the reader
to this paper for more details.
\begin{Def} \label{defSMass Oscillation Property}
The Dirac operator~$\Dir$ has the {\bf{strong mass oscillation property}}
in the interval~$I=(m_L, m_R)$ with domain~$\H^\infty$ if there is a constant~$c>0$ such that
\beq \label{smop}
|\bra \p \psi | \p \phi \ket| \leq c \int_I \, \|\phi_m\|_m\, \|\psi_m\|_m\: dm
\qquad  \textnormal{for all} \: \psi, \phi \in \H^\infty\:.
\eeq
\end{Def}

\noindent The following theorem is proved in~\cite[Theorem~4.2, Proposition~4.3 and
Theorem~4.7]{infinite}.
\begin{Thm} \label{thmSrep}
Assume that the Dirac operator~$\Dir$ has the strong mass oscillation property in the
interval~$I=(m_L, m_R)$. Then there exists a family of linear operators $(\Sig_m)_{m \in I}$
with~$\Sig_m \in \Lin(\H_m)$ which are uniformly bounded,
\[ \sup_{m \in I} \| \Sig_m\| < \infty\:, \]
such that
\beq \label{Smdef}
\bra \p \psi | \p \phi \ket = \int_I (\psi_m \,|\, \Sig_m \,\phi_m)_m\: dm \qquad
\text{for all~$\psi, \phi \in \H^\infty$}\:.
\eeq
The operator~$\Sig_m$ is uniquely determined for every~$m \in I$ by demanding that
for all~$\psi, \phi \in \H^\infty$, the functions~$( \psi_m | \Sig_m \phi_m)_m$ are continuous in~$m$.
Moreover, the operator~$\Sig_m$ is the same for all choices of~$I$ containing~$m$.
Finally, there is a bi-distribution~${\mathcal{P}} \in \D'(\scrM \times \scrM)$ such that the operator~$P$ defined by
\[ P := -\chi_{(-\infty, 0)}(\Sig_m)\, k_m \::\: C^\infty_0(\scrM, S\scrM) \rightarrow \H_m \]
has the representation
\[ \bra \phi | P \psi \ket = {\mathcal{P}}(\overline{\phi} \otimes \psi) \qquad
\text{\textnormal{for all}~$\phi, \psi \in C^\infty_0(\scrM, S\scrM)$} \]
(where~$\overline{\phi} = \phi^\dagger \gamma^0$ is the usual adjoint spinor).
\end{Thm} \noindent
The operator~$P$ is referred to as the {\bf{fermionic projector}}.
We also use the standard notation with an integral kernel,
\begin{align*}
\bra \phi | P \psi \ket &= \iint_{\scrM \times \scrM} \Sl \phi(x) \,|\, {\mathcal{P}}(x,y) \,\psi(y) \Sr_x \: d^4x\: d^4y \\
(P \psi)(x) &= \int_{\scrM} {\mathcal{P}}(x,y) \,\psi(y) \: d^4y \:.
\end{align*}

\section{The Dirac Equation in an Electromagnetic Wave and its Solution} \label{secEMwave}
In the presence of an external electromagnetic field, the potential~$\B$ in the Dirac operator~\eqref{Dirdef}
has the form~$\B = \gamma^j A_j$, where~$A_j$ is the electromagnetic four-potential.
We assume that the electromagnetic potential satisfies the Maxwell equations in the Lorenz gauge, i.e.\
\beq \label{Maxwell}
\Box A^j =0 \qquad \text{with} \qquad \partial_j A^j =0 \:.
\eeq
Moreover, denoting the coordinates of Minkowski space by~$p=(t,x,y,z)$,
we consider the situation where the electromagnetic potential is a plane wave,
\beq \label{Aplane}
A = A(t+x) \:.
\eeq
We point out that we do not assume the wave to be homogeneous or harmonic
(i.e.\ it is not of the form~$A=A_0\, \sin(\omega (t+x)+ \alpha)$).
In particular, our ansatz allows for a description of a laser pulse as mentioned in the introduction.
Clearly, the plane wave ansatz satisfies the first equation in~\eqref{Maxwell}.
The Lorenz gauge condition is
\[ 0 = \partial^j A_j(t+x) = \partial_t A_0(t+x) - \partial_x A_1(t+x) = \big(A_0-A_1)'(t+x) \:, \]
implying that~$A_0 - A_1=:c$ is a real constant.
By a gauge transformation~$A \rightarrow A + \partial \Lambda$ with~$\Lambda = -ct$, we can
arrange that this constant is zero, so that~$A_0=A_1=f(t+x)$.
Performing another gauge transformation with~$\Lambda(t+x)=-\int^{t+x} f(\tau)\, d\tau$
being an indefinite integral of~$f$, we can arrange that~$A_0=A_1=0$ everywhere.
Then the Dirac equation~\eqref{dirac} takes the form
\beq \label{dirac_electro}
\big( i\gamma^j \partial_j + \gamma^2 A_2 + \gamma^3 A_3 - m \big)\, \psi_m =0.
\eeq

As first observed in~\cite{volkow}, the Dirac equation~\eqref{dirac} can be solved by
separation and integration. We now recall this construction in a form suitable
for our purposes (for an alternative method of solving the squared Dirac equation
see~\cite[Section~5.2]{fradkin}). We introduce the null coordinates
\beq \label{sldef}
s = t+x \qquad \text{and} \qquad l = t-x
\eeq
and employ the separation ansatz
\beq \label{ansatz}
\psi_m (t,x,y,z) = e^{-i(k_2 y + k_3 z)} \:\chi(s,l) \qquad
\text{for~$k_2, k_3\in \R$}\:.
\eeq
Then a direct computation using
\begin{align*}
\big( \gamma^0 \partial_t + \gamma^1 \partial_x \big) \chi(s,l)
&= \big( 2 N_+ \partial_s + 2 N_- \partial_l \big) \chi(s,l)  \\
\text{with} \qquad N_\pm \,&\!:= \frac{1}{2}\:\big( \gamma^0 \pm \gamma^1 \big)
\end{align*}
shows that the Dirac equation \eqref{dirac_electro} becomes
\beq \label{dirac_separated}
\bigg( 2i N_+ \partial_s + 2 i N_- \partial_l + \sum_{j=2,3}\gamma^j \big(k_j + A_j(s) \big) - m  \bigg) \chi(s,l)  =0 \:.
\eeq
By the anti-commutation relations~\eqref{anti-com}, one sees that
\beq \label{Nrel}
N_+^2 = 0 = N_-^2 \qquad \text{and} \qquad \{ N_-,N_+\} = \1\:,
\eeq
whereas the operators~$\Pi_\pm$ defined by
\beq \label{Pirel}
\Pi_- := N_- N_+ \hspace{1cm} \text{and} \hspace{1cm} \Pi_+ := N_+ N_-
\eeq
are idempotent and satisfy the relation~$\Pi_- + \Pi_+ = \{ N_-,N_+\} = \1$.
Thus multiplying \eqref{dirac_separated} by $N_+$ from the left and using
the anti-commutation relations yields the equation
\[ \bigg( 2 i N_{+} N_{-}  \partial_l -\sum_{j=2,3} \gamma^j \big(k_j + A_j(s) \big)\, N_+ - N_{+} m  \bigg) \chi(s,l) =0 \:. \]
Multiplying by $N_-$ from the left and using that 
$$N_- N_{+} N_{-} = \{N_-,N_+\} N_- = N_-,$$ 
we write the previous equation as
\beq \label{algebraic_eqn}
-2 i \:\frac{\partial}{\partial l} N_- \chi(s,l) = \bigg( \sum_{j=2,3} \gamma^j \big(k_j + A_j(s) \big)  -  m  \bigg)\,
\Pi_- \chi(s,l)\: .
\eeq
On the other hand, multiplying the Dirac equation \eqref{dirac_separated} by $N_-$ gives
\[ 2 i \frac{\partial}{\partial s} \Pi_- \chi(s,l) = \bigg( \sum_{j=2,3}\gamma^j \big(k_j + A_j(s) \big) + m  \bigg)
N_{-} \chi(s,l) \:. \]
Differentiating with respect to~$l$ and using~\eqref{algebraic_eqn} gives the PDE
\beq \label{PDE1}
4\:\frac{\partial^2}{\partial s \,\partial l} \Pi_- \chi(s,l) = -\bigg( \sum_{j=2,3} \big(k_j + A_j(s) \big)^2 + m^2  \bigg)\, \Pi_{-} \chi(s,l) \:.
\eeq

We now introduce the function~$\zeta(s)$ by
\[ \zeta(s) = \int_0^{s} \bigg( \sum_{j=2,3}\big(k_j + A_j(s')\big)^2 + m^2  \bigg)\: ds' \:. \]
Since
\beq \label{zetapes}
\zeta'(s) = \sum_{j=2,3}\big(k_j + A_j(s)\big)^2 + m^2 \geq m^2 > 0\:,
\eeq
the function~$\zeta$ is strictly monotone increasing and thus a diffeomorphism from~$\R$ to~$\R$.
Therefore, we can take~$\zeta$ as a new null coordinate. We denote the inverse transformation
by~$s = \zeta^{-1} : \R \rightarrow \R$. Using the transformation
\[ \partial_s = \frac{\partial \zeta}{\partial s} \: \partial_\zeta \:, \]
in the new coordinates~$(\zeta, l)$, the PDE \eqref{PDE1} simplifies to
\beq \label{KG1}
4\partial_\zeta \partial_l \,\Pi_- \chi\big(s(\zeta),l \big) = - \Pi_{-} \chi \big(s(\zeta),l \big) \:.
\eeq

In analogy to~\eqref{sldef}, we now introduce variables~$T$ and~$X$ by
\[ \zeta = T+X \qquad \text{and} \qquad l = T-X \:. \]
Then
\begin{gather}
T = \frac{\zeta+l}{2}\:,\qquad X = \frac{\zeta-l}{2} \label{TXdef} \\
\partial_\zeta = \frac{\partial T}{\partial \zeta}\: \partial_T+ \frac{\partial X}{\partial \zeta}\: \partial_X
= \frac{1}{2} \big(  \partial_T+ \partial_X \big) \notag \\
\partial_l = \frac{\partial T}{\partial l}\: \partial_T+ \frac{\partial X}{\partial l}\: \partial_X
= \frac{1}{2} \big(  \partial_T - \partial_X \big) \:. \notag
\end{gather}
As a consequence, rewriting the PDE~\eqref{KG1} in the variables~$T$ and~$X$,
we obtain the two-dimensional Klein-Gordon equation of mass one,
\beq \label{KG2}
\big( \partial_T^2 - \partial_X^2 + 1 \big)\,\Pi_- \chi = 0 \:.
\eeq

\section{Decay in Null Directions} \label{secnull}
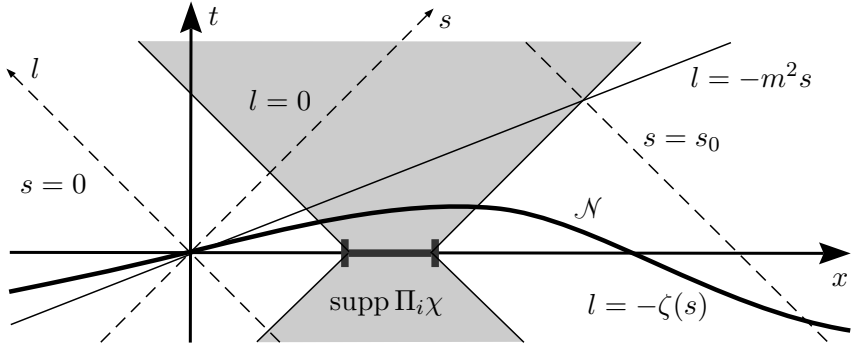
\begin{figure}
\psscalebox{1.0 1.0} 
{
\begin{pspicture}(0,-2.2409408)(13.921123,2.2409408)
\definecolor{colour0}{rgb}{0.8,0.8,0.8}
\definecolor{colour1}{rgb}{0.2,0.2,0.2}
\pspolygon[linecolor=colour0, linewidth=0.02, fillstyle=solid,fillcolor=colour0](4.549456,-1.032106)(1.7344562,1.762894)(8.384457,1.757894)(5.614456,-1.037106)(6.819456,-2.217106)(3.3244562,-2.222106)
\rput[bl](10.941123,-1.4904393){\normalsize{$x$}}
\rput[bl](2.6544561,1.982894){\normalsize{$t$}}
\psline[linecolor=black, linewidth=0.04, arrowsize=0.09300000000000001cm 5.0,arrowlength=1.44,arrowinset=0.39999999999999997]{->}(0.014456177,-1.037106)(11.194456,-1.0504394)
\psline[linecolor=black, linewidth=0.04, arrowsize=0.09300000000000001cm 5.0,arrowlength=1.44,arrowinset=0.39999999999999997]{->}(2.4144561,-2.237106)(2.4244561,2.3028939)
\psline[linecolor=colour1, linewidth=0.1, tbarsize=0.07055555555555557cm 3.0]{|*-|*}(4.469456,-1.047106)(5.669456,-1.047106)
\psline[linecolor=black, linewidth=0.02](5.629456,-1.047106)(6.8494563,-2.2337728)
\psline[linecolor=black, linewidth=0.02](4.524456,-1.027106)(1.7244562,1.772894)
\psline[linecolor=black, linewidth=0.02](4.509456,-1.052106)(3.2994561,-2.227106)
\rput[bl](9.064456,1.132894){\normalsize{$l=-m^2 s$}}
\rput[bl](8.429456,0.31789398){\normalsize{$s=s_0$}}
\rput[bl](7.6961226,-1.9487727){\normalsize{$l = - \zeta(s)$}}
\rput[bl](0.14945617,-0.252106){\normalsize{$s=0$}}
\psline[linecolor=black, linewidth=0.02, linestyle=dashed, dash=0.17638889cm 0.10583334cm, arrowsize=0.05291666666666668cm 2.0,arrowlength=1.4,arrowinset=0.0]{->}(3.6044562,-2.232106)(-0.030543823,1.407894)
\psline[linecolor=black, linewidth=0.02, linestyle=dashed, dash=0.17638889cm 0.10583334cm, arrowsize=0.05291666666666668cm 2.0,arrowlength=1.4,arrowinset=0.0]{<-}(5.649456,2.207894)(1.2244562,-2.232106)
\psline[linecolor=black, linewidth=0.02](8.404456,1.757894)(5.609456,-1.047106)
\psbezier[linecolor=black, linewidth=0.06](0.0061228434,-1.561106)(0.77753663,-1.4004394)(1.1930269,-1.3279338)(2.4351122,-1.0237727)(3.6771977,-0.7196116)(5.1795397,-0.32572928)(6.4760184,-0.45043936)(7.7724977,-0.5751494)(9.389333,-1.7892783)(11.18779,-2.077106)
\psline[linecolor=black, linewidth=0.02, linestyle=dashed, dash=0.17638889cm 0.10583334cm](10.881123,-2.222106)(6.8744564,1.757894)
\psline[linecolor=black, linewidth=0.02](9.601261,1.7537442)(0.027651662,-1.9979563)
\rput[bl](3.194456,0.852894){\normalsize{$l=0$}}
\rput[bl](7.526123,-0.5887727){\normalsize{$\scrN$}}
\rput[bl](4.2694564,-1.8721061){\normalsize{$\text{supp}\, \Pi_i \chi$}}
\rput[bl](0.29445618,1.287894){\normalsize{$l$}}
\rput[bl](5.714456,1.947894){\normalsize{$s$}}
\end{pspicture}
}
\caption{The support of~$\Pi_- \chi$ on the surface~$\scrN = \{l=-\zeta(s)\}$.}
\label{figT}
\end{figure}

In order to formulate the Dirac equation as an evolution equation in lightlike directions,
we need suitable decay properties of solutions in lightlike directions.
In the next lemma, we show that solutions decay even rapidly.
\begin{Lemma} \label{lemmanull}
Let~$\Pi_- \chi(s,l )$ be a solution of the Dirac equation~\eqref{PDE1}
for smooth, compactly supported initial data at~$t=0$,
\beq \label{Cauchy}
\Pi_- \chi|_{t=0}, \;\partial_t \Pi_- \chi|_{t=0} \in C^\infty_0(\R, \C^4) \:.
\eeq
Then the solution decays rapidly in~$l$, uniformly in~$s$, i.e.\ for all~$N>0$ there exists a constant~$c$ such that
\[ \sup_{l \in \R} |l|^N\: \big\|\Pi_- \chi(s,l) \big\| < c \qquad \text{for all~$s \in \R$} \]
(where~$\| . \|$ is any norm on~$\C^4$). The constant~$c$ can be chosen locally uniformly in
the separation constants~$k_2$ and~$k_3$.
\end{Lemma}
\Proof Due to finite propagation speed, the solution of the Cauchy problem with
initial data~\eqref{Cauchy} is in~$\Cisc(\R^{1,1}, \C^4)$.
Let~$\scrN$ be the surface~$\{T=0\}$ (with~$T$ as in~\eqref{TXdef}).
In view of~\eqref{TXdef}, this surface can be
described as a graph over~$s$,
\[ l = - \zeta(s) \:. \]
In particular, the surface is smooth. Moreover, the inequality~\eqref{zetapes} gives the
bound $l'(s) \leq -m^2$ for the slope of this graph (see Figure~\ref{figT}).
Clearly, the straight line $l=-m^2 s$ is a Cauchy surface in~$\R^{1,1}$. Therefore,
the intersection of this straight line with the support of~$\Pi_- \chi$ is compact
and thus contained in the interval~$[-s_0, s_0]$ for sufficiently large~$s_0>0$.
Using that~$\scrN$ lies below this straight line if~$s>0$ and lies above this line if~$s<0$
(see again Figure~\ref{figT}), it follows that also the curve~$\scrN$ does not intersect the support of~$\Pi_- \chi$
for any~$s$ outside this interval, i.e.
\[ \scrN \cap \supp \Pi_- \chi \;\subset\; \big\{ \big(s, - \zeta(s) \big) \text{ with } s \in [-s_0, s_0] \big\}\:. \]
We conclude that the wave function~$\Pi_- \chi$ has compact support on~$\scrN$,
\beq \label{initN}
\Pi_- \chi|_\scrN,\; \partial_T \Pi_- \chi|_\scrN \in C^\infty_0(\scrN, \C^4) \:.
\eeq

We next regard~$\Pi_- \chi$ as the solution of the Cauchy problem of the Klein-Gordon equation~\eqref{KG2}
with initial data~\eqref{initN}. As shown in~\cite[Theorem~7.2.1]{hormanderhyp}, this solution
decays rapidly in null directions (see also the more detailed estimates in~\cite[Section~4.4]{treude}),
in the sense that for all~$N>0$ there exists a constant~$c$ such that
\[ \sup_{l \in \R} |l|^N\: \big\|\Pi_- \chi \big(s(\zeta),l \big) \big\| < c \qquad \text{for all~$\zeta \in \R$} \:. \]
This gives the desired estimate. 
This estimate is locally uniform in~$k_2$ and~$k_3$ because the curve~$\scrN$
as well as the subsequent estimates depend smoothly on these separation constants.
Thus for any compact~$K \subset \R^2$, the constant~$c$ can be chosen uniformly
for all~$(k_2, k_3) \in K$.
\QED

\begin{Lemma} \label{spatial_int_Lemma}
For all~$\psi, \phi \in \H_m$ and~$s \in \R$, the scalar product~\eqref{print}
can be expressed as the following integral over the null hypersurface~$\{s=\text{const}\}$,
\begin{align} \label{spatial_int_null}
(\psi|\phi) &= 2 \pi \int_{\R^3} \Sl \Pi_-\psi | \gamma^0 \,\Pi_-  \phi \Sr_{(s, l,y,z)} \:dl \,dy\, dz \:.
\end{align}
\end{Lemma}
\Proof In view of the polarization identity, we may choose~$\phi= \psi$.
Moreover, using a denseness argument, it suffices to consider a
solution~$\psi$ corresponding to Cauchy data~$\psi_0 = \psi|_{t=0}$ which is smooth and
compactly supported in~$x$, and whose Fourier transform in~$y$ and~$z$ is also compactly supported, i.e.
\[ \hat{\psi}_0(x,k_2, k_3) := \int_{\R^2} \psi_0(x, y,z)\: e^{i k_2 y + i k_3 z}\: dy \: dz \;\in\; C^\infty_0(\R^3, \C^4) \:. \]
Then the solution can be represented as
\[ \psi(t,x,y,z) = \int_{K} e^{-i k_2 y - i k_3 z}\: \chi_{k_2, k_3}(s,l) \: dk_2\: dk_3 \:, \]
where the subset~$K \subset \R^2$ is compact.

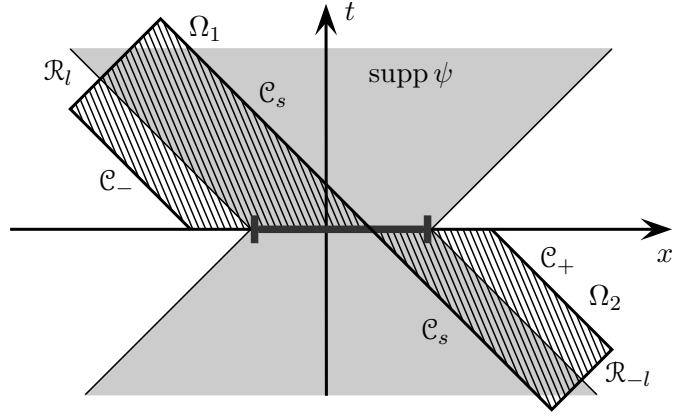
\begin{figure}
{
\begin{pspicture}(0,-2.714142)(13.43,2.714142)
\definecolor{colour0}{rgb}{0.8,0.8,0.8}
\definecolor{colour1}{rgb}{0.2,0.2,0.2}
\pspolygon[linecolor=colour0, linewidth=0.02, fillstyle=solid,fillcolor=colour0](3.2,-0.28585786)(0.8,2.1141422)(8.0,2.1141422)(5.6,-0.28585786)(7.8,-2.485858)(1.0,-2.485858)
\pspolygon[linecolor=colour0, linewidth=0.02, fillstyle=vlines, hatchwidth=0.02, hatchangle=22.5, hatchsep=0.06](4.8,-0.28585786)(6.4,-0.28585786)(8.0,-1.8858578)(7.2,-2.6858578)
\pspolygon[linecolor=colour0, linewidth=0.02, fillstyle=vlines, hatchwidth=0.02, hatchangle=22.5, hatchsep=0.06](2.4,-0.28585786)(0.8,1.3141421)(2.0,2.514142)(4.8,-0.28585786)
\rput[bl](8.6,-0.72585785){\normalsize{$x$}}
\rput[bl](4.45,2.504142){\normalsize{$t$}}
\psline[linecolor=black, linewidth=0.04, arrowsize=0.093cm 5.0,arrowlength=1.44,arrowinset=0.4]{->}(0.0,-0.28585786)(8.8,-0.28585786)
\psline[linecolor=black, linewidth=0.04, arrowsize=0.093cm 5.0,arrowlength=1.44,arrowinset=0.4]{->}(4.205,-2.5008578)(4.2,2.714142)
\psline[linecolor=colour1, linewidth=0.1, tbarsize=0.07055555cm 3.0]{|-|}(3.2,-0.28585786)(5.6,-0.28585786)
\psline[linecolor=black, linewidth=0.02](5.6,-0.28585786)(8.0,2.1141422)
\psline[linecolor=black, linewidth=0.02](5.6,-0.28585786)(7.8,-2.485858)
\psline[linecolor=black, linewidth=0.02](3.2,-0.28585786)(0.8,2.1141422)
\psline[linecolor=black, linewidth=0.02](3.2,-0.28585786)(1.0,-2.485858)
\psline[linecolor=black, linewidth=0.04](2.4,-0.28585786)(0.8,1.3141421)(2.0,2.514142)(7.2,-2.6858578)(8.0,-1.8858578)(6.4,-0.28585786)
\rput[bl](4.77,1.6041422){\normalsize{$\text{supp} \,\psi$}}
\rput[bl](7.91,-2.4058578){\normalsize{$\mathscr{R}_{-l}$}}
\rput[bl](0.45,1.6341422){\normalsize{${\mathscr{R}}_l$}}
\rput[bl](1.175,0.14914215){\normalsize{$\mathscr{C}_-$}}
\rput[bl](7.04,-0.8808578){\normalsize{$\mathscr{C}_+$}}
\rput[bl](5.46,-1.8058579){\normalsize{$\mathscr{C}_s$}}
\rput[bl](3.3,1.3441422){\normalsize{$\mathscr{C}_s$}}
\rput[bl](2.375,2.224142){\normalsize{$\Omega_1$}}
\rput[bl](7.695,-1.3558578){\normalsize{$\Omega_2$}}
\end{pspicture}
}
\caption{Current conservation in a domain with null boundaries.}
\label{fignull}
\end{figure}

We choose an open interval~$J$ such that~$\supp \psi(0, \cdot , y,z) \subset J$ for all~$y,z \in \R$.
We choose the sets~$\Omega_1$ and~$\Omega_2$ as in Figure~\ref{fignull}.
Then, due to Dirac current conservation,
\begin{align}
0 &= \int_{\Omega_1} \partial_j \Sl \psi | \gamma^j \psi \Sr \: d^4x
- \int_{\Omega_2} \partial_j \Sl \psi | \gamma^j \psi \Sr\: d^4x \notag \\
&= \frac{1}{2 \pi}\:(\psi|\psi) - \int_{\mathscr{C}_s} \Sl \psi | \gamma_i \psi \Sr\: d\mu^i_{\mathscr{C}_s}
- \int_{\mathscr{R}_l} \Sl \psi | \gamma_i  \psi \Sr\: d\mu^i_{\mathscr{R}_l}
+ \int_{\mathscr{R}_{-l}} \Sl \psi | \gamma_i \psi \Sr\: d\mu^i_{\mathscr{R}_{-l}} \:, \label{boundary0}
\end{align}
where in the last line we applied the Gauss divergence theorem and
made use of the fact that~$\psi$ vanishes identically on the boundary
components~${\mathscr{C}}_-$ and~${\mathscr{C}}_+$.
If the boundary were space-like, the factors~$d\mu^i$ could be written as
the future-directed unit normal~$\nu^i$ times the volume measure on the boundary surface.
In our setting of null boundaries, the situation is a bit more involved because the normal
is a null vector tangential to the surface, which cannot be normalized.
The correct expressions for the~$d\mu^i$ are obtained most easily
using Fubini and the fundamental theorem of calculus. For example, the 
boundary term on~$\mathscr{C}_s$ is computed by
\begin{align}
0 &= \int_{ \{ t+x \leq s\} } \partial_j \Sl \psi | \gamma^j \psi \Sr \: d^4x 
= \int_{ \{ t+x \leq s\} } \big( \partial_t \Sl \psi | \gamma^0 \psi \Sr +  \partial_x \Sl \psi | \gamma^1 \psi \Sr 
\big)\: d^4x  \notag \\
&= \int_{-\infty}^\infty dy \int_{-\infty}^\infty dz
\int_{-\infty}^\infty dx \:\bigg( \int_{-\infty}^{s-x} \partial_t \Sl \psi | \gamma^0 \psi \Sr\: dt \bigg) \notag \\
&\qquad + \int_{-\infty}^\infty dy \int_{-\infty}^\infty dz
\int_{-\infty}^\infty dt \: \bigg( \int_{-\infty}^{s-t} \partial_x \Sl \psi | \gamma^1 \psi \Sr\: dx \bigg) \notag \\
&= \int_{\R^3} \Sl \psi | \gamma^0 \psi \Sr|_{t=s-x}\: dx \:dy\: dz + \int_{\R^3} \Sl \psi | \gamma^1 \psi \Sr|_{x=s-t}\: dt \:dy\: dz \notag \\
&= \int_{\R^3} \Sl \psi | N_+ \psi \Sr\: dl \:dy \: dz \:, \label{boundary}
\end{align}
showing that~$\gamma_i \,d\mu^i_{\mathscr{C}_s} = N_+\: dl \:dy \: dz$.
The other boundary integrals can be computed similarly.
The boundary term in~\eqref{boundary} can be rewritten as follows,
\beq \begin{split}
\Sl \psi \,|\, N_+  \psi \Sr &=  \Sl \psi \,|\, N_+ N_- N_+  \psi \Sr = \Sl N_+\psi \,|\, \Pi_-  \psi \Sr
= \Sl N_+\psi \,|\, N_- N_+ \Pi_-  \psi \Sr \\
&= \Sl N_- N_+\psi \,|\, N_+ \Pi_-  \psi \Sr
= \Sl \Pi_-\psi \,|\, \gamma^0 \,\Pi_-  \psi \Sr \:,
\end{split} \label{spintransform}
\eeq
where in the last step we used the identity~$N_+ \Pi_- = (\gamma^0 - N_-) \Pi_- = \gamma^0 \,\Pi_-$.
Using this formula in~\eqref{boundary0}, we obtain
\beq \label{psipsi}
\frac{1}{2 \pi}\: (\psi|\psi) = \int_{\mathscr{C}_s} \Sl \Pi_-\psi | \gamma^0 \,\Pi_-  \psi \Sr\: dl \:dy \: dz
+ \int_{\mathscr{R}_l} \Sl \psi | \gamma_i  \psi \Sr\: d\mu^i_{\mathscr{R}_l}
- \int_{\mathscr{R}_{-l}} \Sl \psi | \gamma_i \psi \Sr\: d\mu^i_{\mathscr{R}_{-l}}\:.
\eeq

In view of Lemma~\ref{lemmanull}, the solutions~$\Pi_-\chi_{k_2, k_3}$ all decay rapidly in~$l$,
uniformly for all~$(k_2, k_3) \in K$. This shows that~$\Pi_-  \psi$ decays rapidly in~$l$, uniformly in~$y$ and~$z$.
Therefore, in the limit~$l \rightarrow \infty$ in~\eqref{psipsi}, the integrals over~$\mathscr{R}_{-l}$
and~$\mathscr{R}_l$ vanish, whereas the integral over~${\mathscr{C}}_s$ goes over to the
integral over the null hyperplane~$\{s=\text{const}\}$. This gives the result.
\QED

\section{The Fermionic Signature Operator} \label{secsig}
After these preparations, we are in the position to compute the fermionic signature operator.
It is most convenient to also separate the $l$-dependence with a plane-wave ansatz, i.e.
\beq \label{ansatz2}
\psi (t,x,y,z) = \int_{\R^3} e^{-i k_2 y - i k_3 z}\: e^{-iul} \: \chi_{k_2, k_3,u}(s) \:du\: dk_2\: dk_3 \:.
\eeq
Comparing this ansatz with~\eqref{ansatz}, one sees that~$\chi(s,l) = e^{-iul} \:\chi(s)$.
Using this formula in~\eqref{algebraic_eqn} and~\eqref{PDE1}, the Dirac equation
reduces to one algebraic equation and one ODE,
\begin{align}
2 u N_- \chi(s,l) &= -\bigg( \sum_{j=2,3} \gamma^j \big(k_j + A_j(s) \big)  -  m  \bigg)\, \Pi_- \chi(s,l) 
\label{algebra} \\
4iu\:\frac{d}{d s} \Pi_- \chi_{k_2, k_3,u}(s) &= \bigg( \sum_{j=2,3} \big(k_j + A_j(s) \big)^2 + m^2  \bigg)
\,\Pi_{-} \chi_{k_2, k_3,u}(s) \:. \label{ODE}
\end{align}
Integrating the last equation for initial data on the null surface~$\mathcal{N} := \{s =0\}$, we obtain
the unique solution
\beq \label{dirac_ODE_soln}
\Pi_- \chi_{k_2, k_3,u}(s)
=  \exp \Bigg( -\frac{i}{4u} \int_0^{s} \bigg( \sum_{j=2,3}\big(k_j + A_j(s')\big)^2 + m^2  \bigg)\: ds' \Bigg)\:
\Pi_- \chi_{k_2, k_3,u}(0) \:.
\eeq

We next express the scalar product in terms of the spinor~$\chi_{k_2, k_3,u}(s)$
in the separation ansatz~\eqref{ansatz2}.
\begin{Lemma} \label{lemmasprod}
For any~$s \in \R$, the scalar product can be written as
\[ (\psi|\phi) = (2 \pi)^4 \int_{\R^3} \Sl \Pi_- \chi_{k_2, k_3,u}(s) | \gamma^0 \,\Pi_-  \chi_{k_2, k_3,u}(s) \Sr
\:du \,dk_2\, dk_3 \:. \]
\end{Lemma}
\Proof The formula follows immediately by substituting the ansatz~\eqref{ansatz2} into~\eqref{spatial_int_null}
and applying Plancherel's theorem.
\QED

The remaining task is to compute the space-time inner product in~\eqref{stipMin}.
To this end, we need to consider families of Dirac solution~$(\psi_m)_{m \in I}$
for a varying parameter in an interval~$I = (m_L, m_R)$ with~$0<m_L, m_R$.
We again use the separation ansatz~\eqref{ansatz2}, but now indicate the dependence on the
mass parameter by a superscript~$m$, 
\beq \label{ansatzm}
\psi_m (t,x,y,z) = \int_{\R^3} e^{-i k_2 y - i k_3 z}\: e^{-iul} \: \chi^m_{k_2, k_3,u}(s) \:du\: dk_2\: dk_3 \:.
\eeq
Then, again using Plancherel's theorem,
\begin{align}
\bra \p\psi|\p\phi \ket 
&= \frac{1}{2} \int_{\R^4} \Sl \p\psi| \p \phi\Sr \:  ds \,dl \, dy \, dz \notag \\
&= 4 \pi^3 \int_{\R^4} \Sl \p \chi_{k_2, k_3,u}(s) \,|\, \p \chi_{k_2, k_3,u}(s)\Sr \:  ds \,du \, dk_2 \, dk_3 \notag \\
&= 4 \pi^3 \int_{\R^4} \bigg( \int_I dm \int_I dm' \Sl \chi^m_{k_2, k_3,u}(s) \,|\, \chi^{m'}_{k_2, k_3,u}(s)\Sr 
\bigg) ds \,du \, dk_2 \, dk_3 \:, \label{integrand}
\end{align}
where
\[ \p \chi_{k_2, k_3,u}(s) := \int_I \chi^m_{k_2, k_3,u}(s) \:dm \:. \]

In the next lemma we rewrite the integrand in~\eqref{integrand}
in a more convenient form.
\begin{Lemma} \label{lemma52}
The spinor~$\chi^m_{k_2, k_3,u}(s)$ in the separation ansatz~\eqref{ansatzm} satisfies the relations
\begin{align*}
2u\:\Sl &\chi^m_{k_2, k_3,u}(s) \,|\, \chi^{m'}_{k_2, k_3,u}(s) \Sr \\
&= (m+m')\:  e^{\frac{i}{4u} \big(m^2 - m'^2 \big) s}\:
\Sl \Pi_- \chi^m_{k_2, k_3,u}(0) \,|\, \gamma^0 \,\Pi_- \chi^{m'}_{k_2, k_3,u}(0) \Sr \:.
\end{align*}
\end{Lemma}

We remark for clarity that for~$u=0$, this equation is trivially satisfied because the
spinors~$\Pi_- \chi^m_{k_2, k_3,u}(0)$ and~$\Pi_- \chi^{m'}_{k_2, k_3,u}(0)$
vanish in view of the algebraic equation~\eqref{algebra}, noting that
the matrix~$\sum_{j=2,3} \gamma^j \big(k_j + A_j(s) \big)  -  m$ is invertible
(with its inverse being a multiple of~$\sum_{j=2,3} \gamma^j \big(k_j + A_j(s) \big)  + m$;
see the computation before~\eqref{PDE1}).
\Proof[Proof of Lemma~\ref{lemma52}] For ease of notation, we omit the indices~$k_2$, $k_3$, $u$ as well as the argument~$s$.
Then, similar as in~\eqref{spintransform}, we obtain
\begin{align}
\Sl \chi^m | \chi^{m'} \Sr  
&= \Sl \chi^m \,|\, \Pi_- \chi^{m'} \Sr   + \Sl \chi^m \,|\, \Pi_+ \chi^{m'} \Sr  \notag \\
&= \Sl \Pi_+ \chi^m \,|\, \Pi_- \chi^{m'} \Sr   + \Sl \Pi_- \chi^m \,|\, \Pi_+ \chi^{m'} \Sr \notag \\
&= \Sl\gamma^0 N_- \chi^m |\Pi_- \chi^{m'} \Sr +\Sl \Pi_- \chi^m |\gamma^0 N_- \chi^{m'}\Sr \:.
\label{inter}
\end{align}
Using the abbreviation~$\mathcal{A} \equiv \sum_{j=2,3} \gamma^j(k_j + A_j)$, the
algebraic equation~\eqref{algebra} can be written as
\[ 2u \:N_- \chi^m = -\big( \mathcal{A}-m \big)\, \Pi_- \chi^m \:. \]
Using this equation in~\eqref{inter} gives
\begin{align}
2u\: \Sl \chi^m | \chi^{m'} \Sr  
&= -\Sl\gamma^0 \big( \mathcal{A}-m \big) \Pi_- \chi^m \,|\, \Pi_- \chi^{m'} \Sr 
-\Sl \Pi_- \chi^m \,|\, \gamma^0  \big( \mathcal{A}-m' \big) \Pi_- \chi^{m'}\Sr \notag \\
&= (m+m')  \,\Sl \Pi_- \chi^m \,|\, \gamma^0 \,\Pi_- \chi^{m'} \Sr \:, \label{chimform}
\end{align}
where in the last line we used that~$\mathcal{A}$ is symmetric with respect to the spin scalar product
and anti-commutes with~$\gamma^0$.

We next work out the $s$-dependence. Writing~\eqref{dirac_ODE_soln} in the shorter form
\[ \Pi_- \chi_{k_2, k_3,u}(s)
=  e^{-\frac{i}{4u} \int_0^{s} \big( -\mathcal{A}^2 + m^2  \big)}\:\Pi_- \chi_{k_2, k_3,u}(0) \:, \]
we obtain
\begin{align*}
&\Sl \chi^m_{k_2, k_3,u}(s) \,|\, \chi^{m'}_{k_2, k_3,u}(s) \Sr \\
&= e^{\frac{i}{4u} \int_0^{s} \big( -\mathcal{A}^2 + m^2  \big)}\:
e^{-\frac{i}{4u} \int_0^{s} \big( -\mathcal{A}^2 + m'^2  \big) }\:
\Sl \chi^m_{k_2, k_3,u}(0) \,|\, \chi^{m'}_{k_2, k_3,u}(0) \Sr \\
&= e^{\frac{i}{4u} \big(m^2 - m'^2 \big) s}\:
\Sl \chi^m_{k_2, k_3,u}(0) \,|\, \chi^{m'}_{k_2, k_3,u}(0) \Sr \:.
\end{align*}
Applying~\eqref{chimform} gives the result.
\QED
Using the formula of this lemma in~\eqref{integrand} and formally exchanging the integrals,
one can carry out the $s$-integration using the distributional relation
\[ \int_{-\infty}^\infty e^{\frac{i}{4 u}\: \big(m^2 - m'^2 \big) s}\:ds = 
2\pi \:\delta \Big(\frac{m^2-m'^2}{4u} \Big)\: . \]
In the next lemma we give a rigorous justification of this formal computation.
\begin{Lemma} \label{lemma53}
For any test function~$\eta \in C^\infty_0(I \times I)$,
\[ \int_{-\infty}^\infty ds \int_I dm \int_I dm'\: (m+m')\:  e^{\frac{i}{4u} \big(m^2 - m'^2 \big) s}\: \eta\big(m,m'
\big) = 8 \pi \,|u| \int_I \eta(m,m)\: dm \:. \]
\end{Lemma}
\Proof We insert a convergence-generating factor~$e^{-\varepsilon s^2}$, interchange the integrals,
and carry out the $s$-integration by a Gaussian integral,
\begin{align*}
\int_{-\infty}^\infty & ds \int_I dm \int_I dm'\: (m+m')\:  e^{\frac{i}{4u} \big(m^2 - m'^2 \big) s}\: \eta\big(m,m'
\big) \\
&= \lim_{\varepsilon \searrow 0}
\int_{-\infty}^\infty e^{-\varepsilon s^2} \:ds \int_I dm \int_I dm'\: (m+m')\:
e^{\frac{i}{4u} \big(m^2 - m'^2 \big) s}\: \eta\big(m,m' \big) \\
&= \lim_{\varepsilon \searrow 0}
\int_I dm \int_I dm' \: (m+m')\: \eta\big(m,m' \big)
\int_{-\infty}^\infty e^{-\varepsilon s^2} \:e^{\frac{i}{4u} \big(m^2 - m'^2 \big) s}\:ds \\
&= \lim_{\varepsilon \searrow 0} \int_I dm \int_I dm' \: (m+m')\:  \eta\big(m,m' \big)\:
\sqrt{\frac{\pi}{\varepsilon}}\: \exp \Big(-\frac{(m^2 - m'^2)^2}{64 u^2 \varepsilon} \Big) \:.
\end{align*}
Now the limit~$\varepsilon \rightarrow 0$ can be carried out in the distributional sense to obtain
\begin{align*}
&= \int_I dm \int_I dm' \: (m+m')\: \eta\big(m,m' \big)\: 2\pi \:\delta \Big(\frac{m^2-m'^2}{4u} \Big) \\
&= \int_I dm \int_I dm' \: (m+m')\: \eta\big(m,m' \big)\: 2\pi \:\frac{4 |u|}{m+m'}
\: \delta \big(m-m' \big) \:.
\end{align*}
This concludes the proof.
\QED

Using Lemmas~\ref{lemma52} and~\ref{lemma53} in~\eqref{integrand} 
immediately gives the following result.
\begin{Corollary} \label{corstip}
For all~$\psi, \phi \in \Cisco(\scrM \times I, S\scrM)$,
\begin{align*}
\bra \p\psi|\p\phi \ket 
= (2\pi)^4 \int_I  \bigg( \int_{\R^3} 
\epsilon(u) \;\Sl \Pi_- \chi^m_{k_2, k_3,u}(0) \,|\, \gamma^0 \,\Pi_- \chi^m_{k_2, k_3,u}(0) \Sr
\:du \, dk_2 \, dk_3 \bigg) \: dm \:,
\end{align*}
where $\epsilon(u):=u/|u|$ is the step function.
\end{Corollary}

Comparing the formulas in Corollary~\ref{corstip} and Lemma~\ref{lemmasprod},
one can read off the fermionic signature operators~$\Sig_m$ in~\eqref{Smdef}.
This gives rise to the following result.

\begin{Thm} \label{thmsig}
The Dirac operator~\eqref{Dirdef} in the presence of a smooth plane electromagnetic wave
(i.e.\ with~$\B = \slashed{A}$ and~$A$ according to~\eqref{Maxwell} and~\eqref{Aplane})
has the strong mass oscillation property~\eqref{smop}.
The resulting fermionic signature operators~$(\Sig_m)_{m \in I}$
simply act on~$\H_m$ by multiplication with 
the sign of the separation constant~$u$ in~\eqref{ansatzm}, i.e.
\[ (\Sig_m \psi_m) (t,x,y,z) = \int_{\R^3} e^{-i k_2 y - i k_3 z}\: e^{-iul} \: 
\epsilon(u) \: \chi^m_{k_2, k_3,u}(s) \:du\: dk_2\: dk_3 \:. \]
\end{Thm}
\Proof For families of solutions~$\psi, \phi \in \Cisco(\scrM \times I, S\scrM)$,
one sees immediately that the relation~\eqref{Smdef} holds
for~$\Sig_m$ being the multiplication operators
\beq \label{Sigmex}
\Sig_m \,\chi^m_{k_2, k_3,u} = \epsilon(u)\: \chi^m_{k_2, k_3,u} \:.
\eeq
Since the operators~$\Sig_m$ are obviously bounded by one, it follows that
for all~$\psi, \phi \in \Cisco(\scrM \times I, S\scrM)$ the inequality
\[ \big| \bra \p\psi| \p\phi \ket \big|
\leq \int_I \|\psi_m\|\: \|\phi_m\|\: dm \]
holds. This establishes the strong mass oscillation property for~$\H^\infty$
as in~\eqref{Hinfchoice}. Applying Theorem~\ref{thmSrep}, we obtain the existence of
a unique family of bounded operators~$(\Sig_m)_{m \in I}$ satisfying~\eqref{Smdef}.
Using a denseness argument, one sees that these operators are again given by~\eqref{Sigmex}.
This concludes the proof.
\QED

\section{The Fermionic Projector and Quantum States} \label{secquant}
Exactly as explained in~\cite[Section~3]{finite}, the {\em{fermionic projector}}~$P$
is introduced as the operator
\beq \label{Pdef}
P = -\chi_{(-\infty, 0)}(\Sig_m)\, k_m \::\: C^\infty_0(\scrM, S\scrM) \rightarrow \H_m \:,
\eeq
where~$\Sig$ is the fermionic signature operator,
and~$k_m$ is the {\em{causal fundamental solution}} defined as the difference of the
advanced and retarded Green's operators,
\[ k_m := \frac{1}{2 \pi i} \left( s_m^\vee - s_m^\wedge \right) \::\: C^\infty_0(\scrM, S\scrM) \rightarrow \Cisc(\scrM, S\scrM) \cap \H_m\:. \]
The fermionic projector~$P$ can be represented by a distribution, referred
to as the {\em{kernel of the fermionic projector}}. Namely, just as in~\cite[Section~3.5]{finite},
one shows that there is a unique distribution~${\mathcal{P}} \in \D'(\scrM \times \scrM, \C^{4 \times 4})$ such that
\beq \label{Pkerndef}
\bra \phi | P \psi \ket = {\mathcal{P}} \big( \overline{\phi} \otimes \psi \big)
\qquad \text{for all~$\phi, \psi \in C^\infty_0(\scrM, S\scrM)$}\:.
\eeq

Moreover, applying Araki's construction in~\cite{araki1970quasifree} to the
projection operator~$\chi_{(-\infty, 0)}(\Sig_m)$ gives rise to a distinguished
quasi-free ground state of the second-quantized Dirac field with the property that the
two-point distribution coincides with the kernel of the fermionic projector.
In the language of algebraic QFT, this result is stated as follows:
\begin{Thm} \label{thmstate}
There is an algebra of smeared fields generated by~$\Psi(g)$, $\Psi^*(f)$
together with a pure quasi-free state~$\omega$ with the following properties: \\[0.3em]
(a) The canonical anti-commutation relations hold:
\[ \{\Psi(g),\Psi^*(f)\} = \bra g^* \,|\, k_m\, f \ket \:,\qquad
\{\Psi(g),\Psi(g')\} = 0 = \{\Psi^*(f),\Psi^*(f')\} \:. \]
(b) The two-point distribution of the state is given by
\[ \omega \big( \Psi(g) \,\Psi^*(f) \big) = -\iint_{\scrM \times \scrM} g(x) \,{\mathcal{P}}(x,y) f(y) \: d^4x\, d^4y\:. \]
\end{Thm}
The state~$\omega$ is sometimes referred to as the FP state~\cite{fewster+lang}.
The proof of Theorem~\ref{thmstate} is exactly the same as the proof of~\cite[Theorem~1.4]{hadamard} 
as given in~\cite[Section~6]{hadamard}.

\section{Integral Representation of the Fermionic Projector} \label{secfourier}
In the next lemma, the causal Green's function is separated and expressed in terms
of an ODE in the null coordinate~$s$.
We denote points in Minkowski space by~$q=(s,l,y,z)$ (and similarly with a tilde).

\begin{Lemma} \label{lemmaGreen}
For the Dirac equation~\eqref{dirac_electro}
in the presence of an electromagnetic potential of the form~\eqref{Aplane},
the Green's function can be written as
\beq \label{ansatz3}
s_m \big(q,\tilde{q} \big)
= \int_{\R^3} e^{-i k_2 (y-\tilde{y}) - i k_3 (z-\tilde{z})}\: e^{-iu(l-\tilde{l})} \: s_{k_2, k_3, u}\big(s, \tilde{s} \big)
\:du\: dk_2\: dk_3 \:,
\eeq
where the distribution~$s_{k_2, k_3, u}(s, \tilde{s})$ is of the form
\begin{align}
s_{k_2, k_3, u}\big(s, \tilde{s} \big) &= N_-\: a + \Pi_-\:b
+ \frac{2}{(2 \pi)^3}\: \frac{1}{2u}\: N_+\: \delta \big( s-\tilde{s} \big) \label{s1rel} \\
&\qquad  +\frac{1}{2u}\: \bigg( \sum_{j=2,3} \gamma^j \big(k_j + A_j(s) \big)  +  m  \bigg)\, 
\Big( N_+\: b + \Pi_+\:a \Big)\:, \label{s2rel}
\end{align}
and~$a$ and~$b$ are solutions of the ODEs
\begin{align}
4 i u\: \frac{d a}{d s} &= \bigg( \sum_{j=2,3} \big(k_j + A_j(s) \big)^2 + m^2  \bigg) \:a
+ \frac{4 u}{(2 \pi)^3}\: \delta \big( s-\tilde{s} \big) \label{ODEa} \\
4 i u\: \frac{d b}{d s} &= \bigg( \sum_{j=2,3} \big(k_j + A_j(s) \big)^2 + m^2  \bigg) \:b \notag \\
&\qquad + \bigg( \sum_{j=2,3}\gamma^j \big(k_j + A_j(s) \big) + m  \bigg)
\frac{2}{(2 \pi)^3}\: \delta \big( s-\tilde{s} \big)  \:. \label{ODEb}
\end{align}
\end{Lemma}
\Proof Taking~\eqref{ansatz3} similar to~\eqref{ansatz2} as a separation ansatz,
the distribution~$s_{k_2, k_3, u}$ must satisfy the Dirac equation~\eqref{dirac_separated}
with a $\delta$-distribution as a source term, i.e.\
\beq \label{sdefine}
\bigg( 2i N_+ \partial_s + 2 u\, N_- + \sum_{j=2,3}\gamma^j \big(k_j + A_j(s) \big) - m  \bigg) 
s_{k_2, k_3, u}\big(s, \tilde{s} \big)
= \frac{2}{(2 \pi)^3}\: \delta \big( s-\tilde{s} \big)
\eeq
(the factor of two comes about because~$\delta(t-\tilde{t})\, \delta(x-\tilde{x}) = 2 \,\delta(s-\tilde{s})\, \delta(l-\tilde{l})$).

We now proceed similar as in Section~\ref{secEMwave}: Multiplying the Dirac equation~\eqref{sdefine}
by~$\Pi_-$ and using the relations~\eqref{Nrel} and~\eqref{Pirel}, we obtain the
algebraic equation
\begin{align}
&2 u \, N_- s_{k_2, k_3, u}\big(s, \tilde{s} \big) \notag \\
&=-\bigg( \sum_{j=2,3} \gamma^j \big(k_j + A_j(s) \big)  -  m  \bigg)\, \Pi_- s_{k_2, k_3, u}\big(s, \tilde{s} \big)
+ \frac{2}{(2 \pi)^3}\: \Pi_- \,\delta \big( s-\tilde{s} \big) \:. \label{salgebraic}
\end{align}
Similarly, multiplying~\eqref{sdefine} by~$N_-$ gives the differential equation
\begin{align*}
&2 i \frac{d}{ds} \Pi_- s_{k_2, k_3, u}\big(s, \tilde{s} \big) \\
&=\bigg( \sum_{j=2,3}\gamma^j \big(k_j + A_j(s) \big) + m  \bigg) N_{-} s_{k_2, k_3, u}\big(s, \tilde{s} \big)
+ \frac{2}{(2 \pi)^3}\: N_-\: \delta \big( s-\tilde{s} \big) \:.
\end{align*}
Multiplying by~$2u$ and using~\eqref{salgebraic}, we obtain the ODE
\begin{align}
&4 i u\: \frac{d}{ds} \Pi_- s_{k_2, k_3, u}\big(s, \tilde{s} \big) \notag \\
&=\bigg( \sum_{j=2,3} \big(k_j + A_j(s) \big)^2 + m^2  \bigg) \Pi_{-} s_{k_2, k_3, u}\big(s, \tilde{s} \big) \notag \\
&\qquad +\bigg( \sum_{j=2,3}\gamma^j \big(k_j + A_j(s) \big) + m  \bigg)
\frac{2}{(2 \pi)^3}\: \Pi_- \,\delta \big( s-\tilde{s} \big) + \frac{4 u}{(2 \pi)^3}\: N_-\: \delta \big( s-\tilde{s} \big) \:.
\label{sODE}
\end{align}

In~\eqref{s1rel} and~\eqref{s2rel}, the components involving~$N_\pm$ and~$\Pi_\pm$ are written
separately (note that the Dirac matrices~$\gamma^2$ and~$\gamma^3$ anti-commute with~$N_\pm$
and commute with~$\Pi_\pm$). The ODE~\eqref{sODE} poses conditions for the components
involving~$N_-$ and~$\Pi_-$. By direct computation, one sees that this ODE is equivalent
to~\eqref{ODEa} and~\eqref{ODEb}. Moreover, as is verified by a direct computation,
the algebraic equation~\eqref{salgebraic}
determines the components involving~$N_+$ and~$\Pi_+$ in~\eqref{s1rel} and~\eqref{s2rel}.
This gives the result.
\QED

We point out that the homogeneous part of the ODEs~\eqref{ODEa} and~\eqref{ODEb}
is the ODE~\eqref{ODE}, which can be solved by integrating~\eqref{dirac_ODE_soln}.
The ODEs~\eqref{ODEa} and~\eqref{ODEb} can be solved similarly using the method of variation of constants.
For the retarded Green's functions, one obtains
\begin{align*}
a\big(s,\tilde{s}\big) &= -\frac{i}{(2 \pi)^3} \:\Theta\big(s-\tilde{s} \big) \;e^{ -\frac{i}{4u} \int_{\tilde{s}}^{s}
\big( \sum_{j=2,3} (k_j + A_j)^2 + m^2  \big)} \\
b\big(s,\tilde{s}\big) &= -\frac{i}{4u}\:\frac{2}{(2 \pi)^3}\:\Theta\big(s-\tilde{s} \big)\: \bigg( \sum_{j=2,3}\gamma^j \big(k_j + A_j(\tilde{s}) \big) + m  \bigg) \\
&\qquad \times e^{ -\frac{i}{4u} \int_{\tilde{s}}^{s}
\big( \sum_{j=2,3} (k_j + A_j)^2 + m^2  \big)} \:,
\end{align*}
and similarly for the advanced Green's function. Taking the difference of these Green's functions,
we obtain a corresponding representation of~$k_m$ (see~\eqref{kmdef}).
Using that, according to Theorem~\ref{thmsig}, the factor~$\chi_{(-\infty,0)}(\Sig)$ in~\eqref{Pdef}
amounts to a factor~$\Theta(-u)$, we obtain the following integral representation.

\begin{Thm} \label{thmfourier}
For the Dirac equation~\eqref{dirac_electro}
in the presence of an electromagnetic potential of the form~\eqref{Aplane},
the kernel of the fermionic projector~\eqref{Pkerndef} has the representation
\[ {\mathcal{P}} \big(q,\tilde{q} \big)
= \int_{-\infty}^0 du \int_{\R^2} dk_2\, dk_3\;
e^{-i k_2 (y-\tilde{y}) - i k_3 (z-\tilde{z})}\: e^{-iu(l-\tilde{l})} \: {\mathcal{P}}_{k_2, k_3, u}\big(s, \tilde{s} \big)\:, \]
where the distribution~${\mathcal{P}}_{k_2, k_3, u}(s, \tilde{s})$ is of the form
\beq \begin{split}
&{\mathcal{P}}_{k_2, k_3, u}\big(s, \tilde{s} \big) = N_-\: a\big(s,\tilde{s}\big) + \Pi_-\:b\big(s,\tilde{s}\big) \\
&\qquad  +\frac{1}{2u}\: \bigg( \sum_{j=2,3} \gamma^j \big(k_j + A_j(s) \big)  +  m  \bigg)\, 
\Big( N_+\: b\big(s,\tilde{s}\big) + \Pi_+\:a\big(s,\tilde{s}\big) \Big)\:,
\end{split} \label{Prel}
\eeq
and the functions~$a(s,\tilde{s})$ and~$b(s,\tilde{s})$ are given by
\begin{align}
a\big(s,\tilde{s}\big) &= \frac{1}{(2 \pi)^4} \;e^{ -\frac{i}{4u} \int_{\tilde{s}}^{s}
\big( \sum_{j=2,3} (k_j + A_j(s'))^2 + m^2  \big)\: ds' } \label{aform} \\
b\big(s,\tilde{s}\big) &= \frac{1}{2u\, (2 \pi)^4}\:\bigg( \sum_{j=2,3}\gamma^j \big(k_j + A_j(\tilde{s}) \big) + m  \bigg) 
\nonumber \\
&\qquad \times e^{ -\frac{i}{4u} \int_{\tilde{s}}^{s}
\big( \sum_{j=2,3} (k_j + A_j(s'))^2 + m^2  \big)\: ds' } \:. \label{bform}
\end{align}
\end{Thm}

We remark that the FP state coincides with the ground state
constructed in~\cite[Section~5.2.2]{fradkin}
using a ``frequency splitting'' in the null momentum variable~$u$.

\section{Hadamard Form of the Fermionic Projector} \label{sechadamard}
Recall that the {\em{wave front set}} $\text{WF}\, {\mathcal{P}}$
of a distribution~${\mathcal{P}} \in \D'(\R^n, \C^N)$ is defined
as the complement of the points~$(x, \xi) \in \R^n \times (\R^n \setminus \{0\})$ with the following property:
There exists a test function~$f \in C^\infty_0(\R^n)$ with~$f(x)=1$
and a conical neighborhood~$\Gamma \subset \R^n$ of~$\xi$ such that
\beq \label{WFdef0}
\sup_{\zeta \in \Gamma} \big(1 + |\zeta|)^q \:\big\| \widehat{f\, {\mathcal{P}}}(\zeta) \big\|_{\C^N} < \infty \qquad
\text{for all~$q \in \N_0$}\:.
\eeq
A bi-distribution~${\mathcal{P}} \in \D'(\scrM \times \scrM, \C^{4 \times 4})$
is called {\em{of Hadamard form}} if its wave front set is given by\footnote{Working in a
fixed reference frame, we here identify
the co-tangent space $T^*\scrM$ with~$\scrM \times \R^4$. Moreover,
we implicitly identify tangent and co-tangent vectors using the Minkowski metric.}
\beq \label{WFdef}
\begin{split}
\text{WF}\, {\mathcal{P}} = \Big\{ (x,\xi,y, -\xi) &\in \scrM \times \R^4
\times \scrM \times \R^4 \\
&\:\text{with} \qquad \xi^2 = 0\:, \;\xi^0 < 0 \text{ and } y-x \sim \xi \Big\} \:,
\end{split}
\eeq
where~$\xi^0$ denotes the time component of~$\xi$ in a given reference frame.
Here the notation~$y-x \sim \xi$ means that~$y-x$ is a multiple of the lightlike vector~$\xi$
(more generally, in curved space-time this notation means that
there is a lightlike geodesic from~$x$ to~$y$ whose velocity co-vector equals~$\xi$).
In non-technical terms, the Hadamard form means that the singularities of~${\mathcal{P}}(x,y)$ lie
on the light cone and are formed only of negative frequencies.

This is our main result:
\begin{Thm} \label{thmH}
For the Dirac equation~\eqref{dirac_electro}
in the presence of a smooth electromagnetic potential of the form~\eqref{Aplane},
the kernel of the fermionic projector~${\mathcal{P}}$, \eqref{Pkerndef}, is of Hadamard form.
\end{Thm} \noindent
For the FP state~$\omega$ of Theorem~\ref{thmstate}, this result can be stated as follows:
\begin{Corollary} \label{corH}
The fermionic projector state is a Hadamard state.
\end{Corollary}

The remainder of this section is devoted to the proof of Theorem~\ref{thmH}.
Before beginning, we point out that the methods used in~\cite[Section~3]{light} or~\cite[Section~2.2]{cfs}
for computing the light-cone expansion of the fermionic projector do not apply in our setting
because they rely crucially on decay assumptions on the potential at infinity (see~\cite[Lemma~2.1.2]{cfs}).
In preparation, we write~\eqref{WFdef} independent of a reference frame as
\beq \label{WFP}
\text{WF}\, {\mathcal{P}} = \big\{ (x,\xi,y, -\xi) \:\big|\: \xi^2 = 0\:, \la \mathfrak{n}, \xi \ra \leq 0 \text{ and } y-x \sim \xi \big\} \:,
\eeq
where~$\mathfrak{n}$ is any future-directed timelike vector of~$\scrM$ (and~$\la \mathfrak{n}, \xi \ra$ is the Minkowski
inner product). In our setting, the plane electromagnetic wave distinguishes a lightlike direction.
Namely, writing the plane wave~\eqref{Aplane} for $q\in \scrM$ as   
\[ A(q) = A\big( \la \n, q \ra \big) = A(s)  \qquad \text{with} \qquad \n = (1,-1,0,0) =2 \partial_l \:, \]
the vector~$\n$ describes the direction of the propagation of the electromagnetic wave.
It is future-directed and lightlike, and it is unique up to multiplication by an irrelevant positive constant.
Having this distinguished lightlike direction to our disposal, it is natural to consider the
limiting case that the timelike vector in~\eqref{WFP} goes over to the null vector~$\n$.
In this limiting case, the intersection of the light cone with the hypersurface~$\la \n, \xi \ra =0$
changes from one point to a whole lightlike line,
\[ \big\{ \xi \:\big|\: \xi^2 = 0 \big\} \cap
\big\{ \xi \:\big|\: \la \n, \xi \ra =0 \big\} = \big\{ \lambda \n \:\big|\: \lambda \in \R \big\} \:. \]
In particular, the set in~\eqref{WFP} contains additional vectors~$\xi$ which point to
the future, but which are all collinear to~$\n$.
The delicate point will be to show that these additional vectors do {\em{not}} lie in the
wave front set of~${\mathcal{P}}$.

For clarity, we always denote points in Minkowski
space by latin letters~$p,q,r$, whereas momenta are denoted by Greek letters~$\xi, \eta, \zeta$.
We begin with a preparatory lemma.
\begin{Lemma} \label{lemmarep}
For any test function~$f \in C^\infty_0(\scrM \times \scrM)$,
\begin{align}
\big(\widehat{f\, {\mathcal{P}}} \big)(\zeta,\eta) &= \iint_{\scrM \times \scrM} {\mathcal{P}}(p,q)\, f(p,q)\: e^{-i \zeta p - i \eta q}\: d^4p\, d^4q \label{fourier} \\
&= \Theta\big(-\zeta^s \big) 
\iint_{\scrM \times \scrM} k_m(p, q)\,f(p,q)\:  e^{-i \zeta p - i \eta q}\: d^4p\, d^4q \label{term1} \\
&\qquad +\frac{i}{2 \pi} \:\iint_{\scrM \times \scrM} 
k_m(p,q)\: \hat{g}_{p,q}\Big( \frac{\zeta^s}{2} \Big)\: e^{-i \zeta p - i \eta q}\: d^4p\, d^4q\:, \label{term2}
\end{align}
where~$\zeta^s$ is the s-component of $\zeta$ and, setting $p=(s,l,y,z)$,~$\hat{g}_{p,q}$ is the smooth
function
\beq \label{hatg}
\hat{g}_{p,q}(u) = \int_{-\infty}^\infty d\tilde{l} \:
\int_0^1  d\tau \:\frac{\partial}{\partial l} f\big( (s,l + \tau \tilde{l}, y,z), q \big)
\: e^{-i u \tilde{l}} \:.
\eeq
\end{Lemma}
\Proof Using the result of Theorem~\ref{thmsig} in~\eqref{Pdef}, we obtain
\begin{align*}
{\mathcal{P}}(p,q) &= \int_{\scrM} \big(\chi_{-\infty,0}(\Sig_m)\big)(p,r)\: k_m(r,q)\: d^4r \\
&= \frac{i}{2 \pi} \lim_{\varepsilon \searrow 0} 
\int_{-\infty}^\infty \frac{dl'}{l-l' + i \varepsilon}\: k_m\big((s,l',y,z), q\big)\:,
\end{align*}
where~$\chi_{-\infty,0}(\Sig_m)\big(p,r)$ is the integral kernel of the operator~$\chi_{-\infty,0}(\Sig_m)$,
and the coordinates of~$p$ and~$r$ are denoted by
\[ p=(s,l,y,z) \qquad \text{and} \qquad r=(s,l',y,z) \:. \]
We consider the Fourier integral~\eqref{fourier}.
Using the above relation for~${\mathcal{P}}(p,q)$ gives
\begin{align}
&2\pi i\iint_{\scrM \times \scrM} {\mathcal{P}}(p,q)\, f(p,q)\: e^{-i \zeta p - i \eta q}\: d^4p\, d^4q \notag \\
&=\lim_{\varepsilon \searrow 0} \iint_{\scrM \times \scrM} \int_{-\infty}^\infty \frac{dl'}{l-l' + i \varepsilon}\: k_m\big((s,l',y,z), q\big)
\, f(p,q)\: e^{-i \zeta p - i \eta q}\: d^4p\, d^4q \label{eq0} \\
&=\lim_{\varepsilon \searrow 0} \iint_{\scrM \times \scrM} \int_{-\infty}^\infty \frac{dl'}{l-l' + i \varepsilon}\: k_m(r, q )
\, f(r,q)\: e^{-i \zeta p - i \eta q}\: d^4p\, d^4q \label{eq1} \\
&\quad+\lim_{\varepsilon \searrow 0} \iint_{\scrM \times \scrM} \int_{-\infty}^\infty \frac{dl'}{l-l' + i \varepsilon}\: k_m(r, q)
\,\big(f(p,q) - f(r,q) \big)\:  e^{-i \zeta p - i \eta q}\: d^4p\, d^4q \:. \label{eq2}
\end{align}

In~\eqref{eq1} we combine the integral over~$l'$ with the integrations over~$s,y,z$
to an integral over~$r$, and write the $l$-integral separately,
\[ \eqref{eq1} 
=\lim_{\varepsilon \searrow 0} \iint_{\scrM \times \scrM} \int_{-\infty}^\infty \frac{dl}{l-l' + i \varepsilon}\: k_m\big(r, q\big)
\, f(r,q)\: e^{-i \zeta p - i \eta q}\: d^4r\, d^4q \:. \]
Now the $l$-integration can be carried out using the relations
\begin{align*}
\frac{i}{2 \pi} &\lim_{\varepsilon \searrow 0}  \int_{-\infty}^\infty \frac{dl}{l-l' + i \varepsilon}\: e^{i u l} =
\Theta(-u) \: e^{i u l'} \\
\zeta &= \zeta^s \partial_s + \zeta^l \partial_l + \zeta^y \partial_y + \zeta^z \partial_z \\
\zeta p &= \frac{1}{2}\: \zeta^l s + \frac{1}{2}\: \zeta^s l - \zeta^y y - \zeta^z z \\
\frac{i}{2 \pi} \eqref{eq1} 
&= \Theta\big(-\zeta^s \big) 
\iint_{\scrM \times \scrM} k_m(r, q)\,f(r,q)\:  e^{-i \zeta r - i \eta q}\: d^4r\, d^4q\:.
\end{align*}

In~\eqref{eq2}, on the other hand, we use the relation
\begin{align*}
f(p,q) - f(r,q) &= \int_0^1 \frac{d}{d\tau} f\big( \tau p + (1-\tau) r, q \big)\: d\tau \\
&= \int_0^1 \frac{\partial}{\partial \tilde{l}} 
f\big( (s,\tilde{l}, y,z), q \big)\big|_{\tilde{l} = \tau l + (1-\tau) l'} \: (l-l')\: d\tau\:,
\end{align*}
which we write as
\[ f(p,q) - f(r,q)= g_{r,q}(l-l')\: (l-l') \]
for
\[ g_{r,q}(l-l') :=  \int_0^1 \frac{\partial}{\partial \tilde{l}} 
f\big( (s,\tilde{l}, y,z), q \big)\big|_{\tilde{l} = \tau l + (1-\tau) l'} \:d\tau \:. \]
This makes it possible to take the limit~$\varepsilon \searrow 0$ to obtain
\begin{align*}
\eqref{eq2} &= \iint_{\scrM \times \scrM} \int_{-\infty}^\infty dl'\: k_m(r, q)
\,g_{r,q}(l-l')\:  e^{-i \zeta p - i \eta q}\: d^4p\, d^4q \\
&= \iint_{\scrM \times \scrM} \int_{-\infty}^\infty dl\: k_m(r, q)
\,g_{r,q}(l-l')\:  e^{-i \zeta p - i \eta q}\: d^4r\, d^4q \:.
\end{align*}
Carrying out the $l$-integration by
\[ \int_{-\infty}^\infty dl \:g_{r,q}(l-l')\: e^{-\frac{i}{2}\: \zeta^s (l-l')} = \hat{g}_{r,q}\Big( \frac{\zeta^s}{2} \Big) \:, \]
we obtain
\[ \eqref{eq2} = \iint_{\scrM \times \scrM} k_m(r,q)\: \hat{g}_{r,q}\Big( \frac{\zeta^s}{2} \Big)\: e^{-i \zeta r - i \eta q}
\: d^4r\, d^4q \:. \]
This concludes the proof.
\QED

This lemma makes it possible to prove Theorem~\ref{thmH} provided that the momenta
are not collinear to~$\n$:
\begin{Lemma} \label{lemma1}
The wave front set of the kernel of the fermionic projector has the properties
\begin{align*}
\text{\rm{WF}}\, {\mathcal{P}} &\supset \big\{ (x,\xi,y, -\xi) \:\big|\: \xi^2 = 0\:,\; \la \n, \xi \ra < 0
\text{ and } y-x \sim \xi \big\}  \\
\text{\rm{WF}}\, {\mathcal{P}} &\subset \big\{ (x,\xi,y, -\xi) \:\big|\: \xi^2 = 0\:, \la \n, \xi \ra < 0
\text{ and } y-x \sim \xi \big\} \\
&\qquad \cup \big\{ (x,\xi,y, \eta) \:\big|\: \xi \sim \n \text{ and } \eta \sim \n \big\} \:.
\end{align*}
\end{Lemma}
\Proof Since~${\mathcal{P}}$ satisfies the Dirac equation,
the theorem on the propagation of singularities~\cite[Theorem 5.5]{sahlmann2001microlocal}
implies that the momenta in the wave front set must be characteristic, i.e.
\[ \text{\rm{WF}}\, {\mathcal{P}} \subset \big\{ (x,\xi,y,\eta) \:|\: \xi^2 = \eta^2 = 0 \big\} \:. \]
Moreover, the wave front set of the fundamental solution is given by
\beq \label{WFk}
\text{WF}\, k_m = \big\{ (x,\xi,y, -\xi) \:\big|\: \xi^2 = 0, \xi \neq 0 \text{ and } y-x \sim \xi \big\} \:.
\eeq
This can be seen in various ways: One method is to adapt the methods in~\cite[Section~4]{baer+fredenhagen}
to the vector-valued situation; see also~\cite{sahlmann2001microlocal}.
Alternatively, the light-cone expansion (see~\cite[Section~2]{light} or the textbook~\cite[\S2.2.2]{cfs})
gives a method of determining the singularity structure of~$k_m(x,y)$ explicitly.
This analysis shows that~$k_m(x,y)$ is smooth away from the light cone (i.e.\ if~$(y-x)^2 \neq 0$),
and has singularities on the light cone which give rise precisely to the wave front set~\eqref{WFk}.

In view of the symmetry of the kernel of the fermionic projector~${\mathcal{P}}(p,q)^* = {\mathcal{P}}(q,p)$,
it suffices to consider the case that~$\xi$ is not collinear to~$\n$, i.e.   
\[\xi^s \equiv \la \xi, \n \ra \neq 0 \:. \]
Being the Fourier transform of a smooth function, the function~$\hat{g}_{p,q}(u)$ in~\eqref{hatg}
decays rapidly in~$u$. As a consequence, the summand~\eqref{term2} decays rapidly for all~$\tilde{\zeta}$
in a cone around~$\xi$, uniformly in~$\eta$. Therefore, the summand~\eqref{term2} does not contribute
to the wave front set.
The remaining summand~\eqref{term1} shows that the point~$(x,\xi, y, \eta)$ is in the wave front set
of~${\mathcal{P}}$ if and only if it is in the wave front set of~$k_m$
and~$\xi^s<0$. This concludes the proof.
\QED
In view of this lemma, it remains to consider the case that~$\xi \sim \n$ and~$\eta \sim \n$.
In this case, the representation of Lemma~\ref{lemmarep} is not useful, mainly because
the function~$\hat{g}_{p,q}$ in~\eqref{term2} is not compactly supported in~$p$ and~$q$
as required in the definition of the wave front set~\eqref{WFdef0}. We proceed in several steps:

\begin{Lemma} \label{lemma2}
Assume that the vector~$y-x$ is not collinear to~$\n$. Then~$(x,\xi, y, \eta)$
is not in the wave front set of~${\mathcal{P}}$ for all~$\xi, \eta \in \R^4 \setminus \{0\}$.
\end{Lemma}
\Proof We choose a function~$\Theta \in C^\infty(\R)$ with
\[ \Theta(l) = \frac{1}{l} \qquad \text{for all~$l \in \R \setminus [-1,1]$}\:. \]
Then, setting again $r=(s,l',y,z)$, we may rewrite~\eqref{eq0} as
\begin{align}
& 2\pi i \iint_{\scrM \times \scrM} {\mathcal{P}}(p,q)\, f(p,q)\: e^{-i \zeta p - i \eta q}\: d^4p\, d^4q \notag \\
&= \lim_{\varepsilon \searrow 0}\iint_{\scrM \times \scrM} \int_{-\infty}^\infty \frac{dl}{l-l' + i \varepsilon}\: k_m(r, q)
\, f(p,q)\: e^{-i \zeta p - i \eta q}\: d^4r\, d^4q \notag \\
&= \lim_{\varepsilon \searrow 0}\iint_{\scrM \times \scrM} \int_{-\infty}^\infty dl \bigg( \frac{1}{l-l' + i \varepsilon} - \Theta(l-l') \bigg) k_m(r, q)
 f(p,q)\: e^{-i \zeta p - i \eta q}\: d^4r\, d^4q \label{t1} \\
&\qquad+\lim_{\varepsilon \searrow 0} \iint_{\scrM \times \scrM} \int_{-\infty}^\infty dl\: \Theta(l-l')\: k_m(r, q)
\, f(p,q)\: e^{-i \zeta p - i \eta q}\: d^4r\, d^4q \:. \label{t2}
\end{align}
Note that in~\eqref{t1} the integrand vanishes if~$|l-l'| >1$. Moreover, since~$f$ has compact
support, the integrand vanishes if~$|l|>L$ for some sufficiently large~$L>0$ (uniformly in~$p$ and~$q$).
Therefore, the integrand vanishes if~$|l'| > L+1$. Thus, choosing a cutoff function~$\chi_L \in C^\infty_0(\R)$
with~$\chi_L|_{[-L-1, L+1]} \equiv 1$,
\[ \eqref{t1} =
\lim_{\varepsilon \searrow 0} \iint_{\scrM^2 } \int_{-\infty}^\infty dl\: \bigg( \frac{1}{l-l' + i \varepsilon} - \Theta(l-l') \bigg)\: \chi_L(l')\:
k_m(r, q)\: f(p,q)\: e^{-i \zeta p - i \eta q}\: d^4r\, d^4q \:. \]
Next, we expand the function~$f$ in the integrand as a Taylor polynomial in~$l$ around~$l'$, i.e.\
\[ f(p,q) = \sum_{k=0}^K \frac{\partial^k}{\partial l'^k} f(r,q)\: (l-l')^k + R_K(p,q,l-l') \:, \]
where the error term~$R_K$ is of the order~$o((l-l')^K)$,
is smooth in~$p$ and~$q$ and has compact support in~$q$.
For the Taylor polynomial, we can carry out the $l$-integrals term by term,
\[ \lim_{\varepsilon \searrow 0} \int_{-\infty}^\infty dl\: \bigg( \frac{1}{l-l' + i \varepsilon} - \Theta(l-l') \bigg)
\: e^{i u l}\:\frac{\partial^k}{\partial l'^k} f(r,q)\: (l-l')^k
= g_k(u)\: e^{iul'} \:  \frac{\partial^k f(r,q)}{\partial l'^k}  \:, \]
where the function
\[ g_k(u) := \lim_{\varepsilon \searrow 0}
\int_{-\infty}^\infty \frac{1}{(l+ i \varepsilon)^{1-k}}\: \: e^{iul} \: dl \]
is obviously smooth and bounded. We thus obtain
\begin{align}
&\iint_{\scrM \times \scrM} {\mathcal{P}}(p,q)\, f(p,q)\: e^{-i \zeta p - i \eta q}\: d^4p\, d^4q \notag \\
&= \sum_{k=0}^K g_k(u)
\iint_{\scrM \times \scrM} k_m(r, q)\: \frac{\partial^k f(r,q)}{\partial l'^k}\: \chi_L(l')
\: e^{-i \zeta r - i \eta q}\: d^4r\, d^4q \label{term4} \\
&\qquad + \iint_{\scrM \times \scrM} \int_{-L-1}^{L+1} dl'\: \bigg( \frac{1}{l-l' + i \varepsilon} - \Theta(l-l')
\bigg)\notag \\
&\qquad \qquad \times\:R_K(p,q,l-l') \: \chi_L(l')\: k_m\big((s,l',y,z), q\big)
\: e^{-i \zeta p - i \eta q}\: d^4p\, d^4q \label{term5} \\
&\qquad+ \iint_{\scrM \times \scrM} \int_{-\infty}^\infty dl'\: \Theta(l-l')\:  k_m\big((s,l',y,z), q\big)
\, f(p,q)\: e^{-i \zeta p - i \eta q}\: d^4p\, d^4q \:. \label{term6}
\end{align}

Now we can argue as follows: Since~$(x,\xi, y, \eta)$ is not in the wave front set of~$k_m$,
the contribution~\eqref{term4} clearly does not contribute to the wave front set.
Next, as the vector~$y-x$ is not collinear to~$\n$,
the straight line~$\{ (s,l', y,z) \:|\: l' \in \R\}$ intersects the light cone transversely.
As a consequence, the convolution in~\eqref{term6}
\[ \int_{-\infty}^\infty dl'\: \Theta(l-l')\:  k_m\big((s,l',y,z), q\big) \]
gives rise to a smooth function in~$p$ and~$q$, implying that its Fourier transform
decays rapidly in~$\zeta$ and~$\eta$. Hence~\eqref{term6} does not contribute to the
wave front set. 

Finally, the term~\eqref{term5} again involves a convolution,
\beq \label{conv}
\int_{-L-1}^{L+1} dl'\: \bigg( \frac{1}{l-l' + i \varepsilon} - \Theta(l-l')
\bigg) \:R_K(p,q,l-l') \: \chi_L(l')\: k_m\big((s,l',y,z), q\big) \:.
\eeq
Since~$R_K$ vanishes at~$l=l'$, this factor compensates
the pole of the first factor in the integrand. Even more, by choosing~$K$ sufficiently
large, we can make the convolution kernel arbitrarily smooth. More precisely,
\[ \bigg( \frac{1}{\cdot + i \varepsilon} - \Theta(\cdot)
\bigg) \:R_K(p,q, \cdot )  \in C^{K-2}(\R) \:. \]
As worked out in detail in~\cite{light}, the leading singularities of~$k_m$ on the light cone
are of the form~$\sim \slashed{\xi} \, \delta'((y-x)^2)$ and~$\sim \text{PP}/((y-x)^4)$.
As a consequence, the convolution integral~\eqref{conv} has the regularity~$C^{K-4}$ in~$p$ and~$q$.
Since~$K$ is arbitrary, the Fourier transform of~\eqref{conv} decays rapidly in~$\zeta$ and~$\eta$.
This concludes the proof.
\QED

\begin{Lemma} \label{lemma3}
Assume that~$\xi, \tilde{\eta} \in \R^4 \setminus \{0\}$ and~$\tilde{\eta} \neq -\xi$.    
Then the point~$(x,\xi, y, \tilde{\eta})$ is not in the wave front set of~${\mathcal{P}}$ for all~$x,y \in \scrM$.
\end{Lemma}
\Proof We choose a conical set~$\Gamma$ around~$(\xi, \tilde{\eta})$ such that~$\eta \neq -\zeta$
for all~$(\zeta, \eta) \in \Gamma$. We choose an index~$j$ such that~$\eta^j \neq -\zeta^j$. Then
\begin{align}
& \big(-i (\eta^j + \zeta^j)\big)^K \iint_{\scrM \times \scrM} {\mathcal{P}}(p,q)\, f(p,q)\:
e^{-i \zeta p - i \eta q} \: d^4p\, d^4q \notag \\
&=\iint_{\scrM \times \scrM} {\mathcal{P}}(p,q)\, f(p,q)\: \bigg( 
\Big( \frac{\partial}{\partial p^j} + \frac{\partial}{\partial q^j} \Big)^K e^{-i \zeta p - i \eta q} \bigg) \: d^4p\, d^4q \notag \\
&=(-1)^K \iint_{\scrM \times \scrM} \bigg( \Big( \frac{\partial}{\partial p^j} + \frac{\partial}{\partial q^j} \Big)^K
{\mathcal{P}}(p,q)\, f(p,q) \bigg) \: e^{-i \zeta p - i \eta q} \: d^4p\, d^4q \:. \label{diffP}
\end{align}
As is obvious from the formulas of the light cone expansion in~\cite{light},
the singular terms in the light cone expansion depend only on the difference vector~$p-q$.
Thus their derivatives cancel in the combination~$\partial_p + \partial_q$.
Hence in~\eqref{diffP} only the electromagnetic potential in the light cone expansion is differentiated.
As a consequence, the derivatives in~\eqref{diffP} do not increase the order of the singularities
on the light cone. Therefore, \eqref{diffP} shows that the Fourier integral in~\eqref{fourier}
decays rapidly in~$\zeta$ and~$\eta$. This concludes the proof.
\QED

\begin{Lemma} \label{lemma4}
Assume that~$\xi \sim \n$, $\xi^0>0$ and~$x-y \sim \xi$.
Then the point~$(x,\xi, y, -\xi)$ is not in the wave front set of~${\mathcal{P}}$ for all~$x,y \in \scrM$.
Conversely, if~$\xi \sim \n$, $\xi^0<0$ and~$x-y \sim \xi$, then
the point~$(x,\xi, y, -\xi)$ lies in the wave front set of~${\mathcal{P}}$.
\end{Lemma}
\Proof We write the kernel~${\mathcal{P}}_{k_2, k_3, u}$ in~\eqref{Prel} as
\[ {\mathcal{P}}_{k_2, k_3, u}\big(s, \tilde{s} \big) = g_{k_2, k_3, u}(s)\: e^{ -\frac{i}{4u} \int_{\tilde{s}}^{s}
\big( \sum_{j=2,3} (k_j + A_j(s'))^2 + m^2  \big)\: ds' } \:, \]
where~$g_{k_2, k_3, u}(s)$ is a smooth function in all variables~$k_2$, $k_3$, $u \neq 0$ and~$s$.
We now set~$\tilde{s}=0$, multiply by a non-trivial test function~$f(s)$ and
compute the Fourier transform,
\beq \label{Fdef}
F_{k_2, k_3, u}(v) := \int_{-\infty}^\infty f(s)\: g_{k_2, k_3, u}(s)\: e^{ -\frac{i}{4u} \int_0^{s}
\big( \sum_{j=2,3} (k_j + A_j(s'))^2 + m^2  \big)\: ds' } \: e^{i v s}\: ds \:.
\eeq

We first consider the case
\beq \label{vrange}
v > \frac{m^2}{8u}\:.
\eeq
Then the function~$h(s)$ defined by
\[ h(s) = v s - \frac{1}{4u}\: \int_0^{s}
\Big( \sum_{j=2,3} (k_j + A_j(s'))^2 + m^2  \Big)\: ds' \]
is strictly increasing, because
\[ h'(s) = v - \frac{1}{4u}\: \Big( \sum_{j=2,3} (k_j + A_j(s))^2 + m^2  \Big) 
> -\frac{m^2}{8u} > 0\:. \]
Therefore, we can introduce~$h$ as a new integration variable to obtain
\[ F_{k_2, k_3, u}(v) = \int_{-\infty}^\infty \eta\big(s(h) \big)\: e^{i h}\: \frac{dh}{h'(s)} \\
= \int_{-\infty}^\infty \frac{\eta\big(s(h) \big)}{h'(s)}\:  e^{i h}\: dh \:, \]
where we used the abbreviation~$\eta = f\: g_{k_2, k_3, u}$.
Now one can integrate by parts,
\begin{align}
F_{k_2, k_3, u}(v) &= \int_{-\infty}^\infty \frac{\eta\big(s(h) \big)}{h'(s)}\:  \Big( -i \frac{d}{dh} e^{i h} \Big)\: dh
= i \int_{-\infty}^\infty \Big( \frac{d}{dh} \frac{\eta\big(s(h) \big)}{h' \big(s(h) \big)} \Big)\:  e^{i h}\: dh \notag \\
&= i \int_{-\infty}^\infty \Big( \frac{\eta'(s)}{h'(s)^2} - \frac{h''(s)\: \eta(s)}{h'(s)^3} \Big)\:  e^{i h}\: dh \:. \label{intpart}
\end{align}
Iteration shows that~$F_{k_2, k_3, u}(v)$ decays rapidly as~$v \rightarrow \infty$
(note that~$h'(s)$ grows linearly in~$v$, whereas~$h''(s)$ and~$\eta(s)$ are independent of~$v$).
As this estimate is uniform for~$k_2, k_3$ and~$v$ in a conical neighborhood
pointing to the future,
it follows that~$(x,\xi, y, -\xi)$ is not in the wave front set if~$\xi \sim \n$ and~$\xi^0>0$.

In order to prove that the point~$(x,\xi, y, -\xi)$ is in the wave front set if~$\xi \sim \n$ and~$\xi^0<0$,
we proceed indirectly and assume that this point is not in the wave front set.
Then the function~$F_{k_2, k_3, u}(v)$ decays rapidly as~$v \rightarrow -\infty$, locally uniformly
in~$u$, $k_2$ and~$k_3$. On the other hand, the integration-by-parts method~\eqref{intpart}
gives upper bounds for~$F_{k_2, k_3, u}$ in the range~\eqref{vrange}. Combining these results, we conclude that
\[ \lim_{u \nearrow 0} F_{k_2, k_3, u} = 0 \quad \text{in~$L^2(\R)$} \]
for all~$k_2$ and~$k_3$. On the other hand, applying Plancherel's theorem to~\eqref{Fdef}, we obtain
\[ \big\| F_{k_2, k_3, u} \big\|_{L^2(\R)}^2 = 2 \pi 
\int_{-\infty}^\infty \big| f(s)\: g_{k_2, k_3, u}(s) \big|^2\: ds\:, \]
which is bounded away from zero as~$u$ tends to zero (note that~$g_{k_2, k_3, u}$
is non-zero in this limit in view of the explicit representation in~\eqref{Prel}).
This is a contradiction. Hence the point~$(x,\xi, y, -\xi)$ must be in the wave front set.
This concludes the proof.
\QED

Theorem~\ref{thmH} follows immediately by combining the results of Lemmas~\ref{lemma1}--\ref{lemma4}.

\section{Example: A Harmonic Plane Electromagnetic Wave} \label{secexharmonic}
In this section we compute and discuss the kernel of the fermionic projector
in closed form for a harmonic plane electromagnetic wave.
Thus we choose the electromagnetic potential~$A$ in~\eqref{Aplane} and~\eqref{dirac_electro} as
\beq \label{Aharmonic}
A_2 = \lambda \,\cos \big( \Omega \,(t+x) \big) \:,\qquad A_3 = 0
\eeq
for real parameters~$\Omega \neq 0$ and~$\lambda$. For this potential,
the integrals in~\eqref{aform} and~\eqref{bform} can be carried out to obtain
\begin{align*}
&\int_{\tilde{s}}^{s} \bigg( \sum_{j=2,3}\big(k_j + A_j(s')\big)^2 + m^2  \bigg)\: ds' \\
&= \int_{\tilde{s}}^{s} \bigg( \Big( k_2 +2 + \lambda \,\cos ( \Omega s' ) \Big)^2 + k_3^2 + m^2  \bigg)\: ds' \\
&= \int_{\tilde{s}}^{s} \bigg( k_2^2 + 2 k_2 \,\lambda \,\cos (\Omega s') + \lambda^2 \cos^2 (\Omega s') + k_3^2 + m^2  \bigg)\: ds' \\
&= \big( k_2^2 + k_3^2 + m^2 \big) \,s' + \frac{2 k_2 \,\lambda}{\Omega}\: \sin(\Omega s') 
+ \frac{\lambda^2 s'}{2} + \frac{\lambda^2}{4 \Omega}\: \sin(2 \Omega s') \Big|_{s'=\tilde{s}}^{s'=s} \\
&= \Big( k_2^2 + k_3^2 +\frac{\lambda}{2} + m^2 \Big) \,\big(s-\tilde{s} \big) \\
&\qquad + \frac{2 k_2 \,\lambda}{\Omega}\: \Big( \sin(\Omega s) - \sin(\Omega \tilde{s}) \Big)
+ \frac{\lambda^2}{4 \Omega}\: \Big( \sin(2 \Omega s) - \sin(2 \Omega \tilde{s}) \Big) \:.
\end{align*}
Hence the exponentials in~\eqref{aform} and~\eqref{bform} become
\begin{align}
&\exp \Bigg( -\frac{i}{4u} \int_{\tilde{s}}^{s} \bigg( \sum_{j=2,3}\big(k_j + A_j(s)\big)^2 + m^2  \bigg)\: ds \Bigg) \notag \\
&= \exp \bigg( -\frac{i}{4u} \Big( k_2^2 + k_3^2 + \frac{\lambda^2}{2} + m^2 \Big) \,\big(s-\tilde{s} \big) \bigg) \\
&\qquad \times \bigg( \sum_{n=0}^\infty \frac{1}{n!} \Big(-\frac{i k_2 \,\lambda}{2 \Omega u}\: \big( \sin(\Omega s)
- \sin(\Omega \tilde{s}) \big) \Big)^n \bigg) \label{fact1} \\
&\qquad \times \bigg( \sum_{n'=0}^\infty \frac{1}{n'!} \Big( -\frac{i \lambda^2}{16 \Omega u}\: \big( \sin(2 \Omega s)
- \sin(2 \Omega \tilde{s}) \big) \Big)^{n'} \bigg) \:. \label{fact2}
\end{align}

Let us explain what these formulas mean. To this end, we fix~$\tilde{q}$ and take the Fourier transform
of~$k_m(q,\tilde{q})$ in the variable~$q$. We again denote the momenta corresponding to~$l$, $y$ and~$z$
by~$u$, $k_2$, and~$k_3$, respectively. 
Then, according to~\eqref{Sigmex}, the factor~$\chi_{-(\infty, 0)}(\Sig_m)$ in~\eqref{Pdef} simply gives rise to a
factor~$-\Theta(-u)$. 
Moreover, we denote the momentum corresponding to~$s$ by~$v$, so that
\[ u = \frac{1}{2} \:\big( \omega + k_1 \big) \:,\qquad v = \frac{1}{2} \:\big( \omega - k_1 \big) \:. \]
Expanding the trigonometric functions in~\eqref{fact1} and~\eqref{fact2} into
plane waves, one sees that the momenta~$v$ take the discrete values
\beq \label{discrete}
v \in \frac{1}{4u} \Big( k_2^2 + k_3^2 + \frac{\lambda^2}{2} + m^2 \Big)
+ \Omega \,\Z \:.
\eeq
In the case~$\lambda=0$ when the amplitude of the electromagnetic wave is zero,
the first summand gives the usual dispersion relation~$4 uv = k_2^2 + k_3^2 + m^2$
of the plane-wave solutions of the Dirac equation. This dispersion relation is modified
by the plane wave, effectively changing the mass.
More importantly, in~\eqref{discrete} one gets additional momenta
where~$v$ is modified by multiples of~$\Omega$. These additional contributions can be
understood from the fact that the plane wave mixes different momenta.

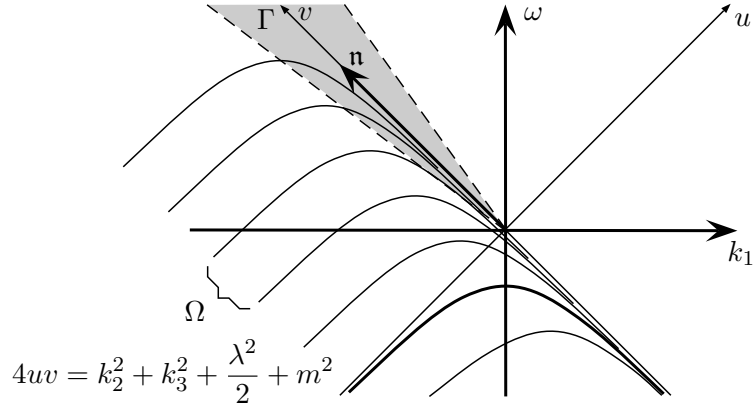
\begin{figure}
{
\begin{pspicture}(0,-2.6625078)(9.85,2.6625078)
\definecolor{colour0}{rgb}{0.8,0.8,0.8}
\pspolygon[linecolor=white, linewidth=0.02, fillstyle=solid,fillcolor=colour0](2.58,2.6474922)(6.585,-0.38250786)(4.405,2.652492)
\rput[bl](9.51,-0.82750785){\normalsize{$k_1$}}
\rput[bl](6.805,2.4324923){\normalsize{$\omega$}}
\psline[linecolor=black, linewidth=0.04, arrowsize=0.093cm 5.0,arrowlength=1.44,arrowinset=0.4]{->}(2.36,-0.35750785)(9.64,-0.36750785)
\psline[linecolor=black, linewidth=0.04, arrowsize=0.093cm 5.0,arrowlength=1.44,arrowinset=0.4]{->}(6.56,-2.5725079)(6.555,2.642492)
\psline[linecolor=black, linewidth=0.02, arrowsize=0.05291667cm 2.0,arrowlength=1.4,arrowinset=0.0]{->}(4.355,-2.5575078)(9.55,2.652492)
\rput[bl](4.475,1.8424921){\normalsize{$\mathfrak{n}$}}
\rput[bl](9.6,2.3574922){\normalsize{$u$}}
\psline[linecolor=black, linewidth=0.02, arrowsize=0.05291667cm 2.0,arrowlength=1.4,arrowinset=0.0]{->}(8.755,-2.5575078)(3.56,2.6374922)
\rput[bl](3.795,2.4424922){\normalsize{$v$}}
\psbezier[linecolor=black, linewidth=0.04](4.475,-2.5175078)(5.0434804,-1.9831059)(5.5710015,-1.4515477)(6.0050783,-1.2475078582763672)(6.4391546,-1.0434679)(6.7861724,-1.0401828)(7.275482,-1.3541232)(7.764792,-1.6680636)(8.099817,-2.0036218)(8.66,-2.5381231)
\psbezier[linecolor=black, linewidth=0.02](3.875,-1.9175079)(4.44348,-1.383106)(4.9710016,-0.8515478)(5.4050784,-0.6475078582763671)(5.8391547,-0.44346792)(6.1861725,-0.44018286)(6.6754823,-0.7541233)(7.164792,-1.0680636)(7.499817,-1.4036218)(8.06,-1.9381232)
\psbezier[linecolor=black, linewidth=0.02](3.275,-1.3175079)(3.84348,-0.78310597)(4.3710017,-0.25154778)(4.805078,-0.047507858276367186)(5.239155,0.15653208)(5.5861726,0.15981713)(6.0754824,-0.15412325)(6.564792,-0.46806362)(6.899817,-0.80362177)(7.46,-1.3381232)
\psbezier[linecolor=black, linewidth=0.02](2.675,-0.71750784)(3.2434802,-0.18310595)(3.7710016,0.3484522)(4.205078,0.5524921417236328)(4.639155,0.7565321)(4.986172,0.7598171)(5.4754825,0.44587675)(5.9647923,0.13193639)(6.299817,-0.20362177)(6.86,-0.73812324)
\psbezier[linecolor=black, linewidth=0.02](5.545,-2.5725079)(5.90348,-2.298106)(6.1710014,-2.0515478)(6.605078,-1.847507858276367)(7.0391545,-1.6434679)(7.3861723,-1.6401829)(7.875482,-1.9541233)(8.364792,-2.2680635)(8.504817,-2.4436219)(8.635,-2.5581234)
\psbezier[linecolor=black, linewidth=0.02](2.075,-0.11750786)(2.6434803,0.41689405)(3.1710017,0.94845223)(3.6050782,1.152492141723633)(4.0391545,1.3565321)(4.3861723,1.3598171)(4.875482,1.0458767)(5.364792,0.7319364)(5.699817,0.39637822)(6.26,-0.13812324)
\psbezier[linecolor=black, linewidth=0.02](1.475,0.48249215)(2.0434802,1.0168941)(2.5710015,1.5484523)(3.005078,1.7524921417236328)(3.4391549,1.9565321)(3.7861724,1.9598172)(4.275482,1.6458768)(4.764792,1.3319364)(5.099817,0.99637824)(5.66,0.46187675)
\rput[bl](0.0,-2.6625078){$\displaystyle{4 uv = k_2^2 + k_3^2 + \frac{\lambda^2}{2} + m^2}$}
\psline[linecolor=black, linewidth=0.04, arrowsize=0.093cm 5.0,arrowlength=1.44,arrowinset=0.4]{->}(6.5452833,-0.3417556)(4.3497167,1.8467399)
\psline[linecolor=black, linewidth=0.02, linestyle=dashed, dash=0.17638889cm 0.10583334cm](6.565,-0.37250787)(2.59,2.6374922)
\psline[linecolor=black, linewidth=0.02, linestyle=dashed, dash=0.17638889cm 0.10583334cm](6.565,-0.35750785)(4.415,2.642492)
\rput[bl](3.255,2.3174922){\normalsize{$\Gamma$}}
\psline[linecolor=black, linewidth=0.02](2.5971572,-0.8396652)(2.5971572,-0.9810865)(2.7385786,-1.1225078)(2.7385786,-1.2639292)(2.88,-1.2639292)(3.0214214,-1.4053506)(3.1628428,-1.4053506)
\rput[bl](2.29,-1.5675079){\normalsize{$\Omega$}}
\end{pspicture}
}
\caption{The Fourier transform of~${\mathcal{P}}(.,\tilde{q})$ for a harmonic plane electromagnetic wave.}
\label{figharmonic}
\end{figure}

The resulting situation is depicted in Figure~\ref{figharmonic}.
One sees that~${\mathcal{P}}$ does involve contributions of arbitrarily large positive frequency.
However, due to the factorials~$n!$ and~$n'!$ in~\eqref{fact1} and~\eqref{fact2}, these contributions
decay rapidly for large~$v$. This is the reason why the bi-distribution~${\mathcal{P}}(q,\tilde{q})$
is of Hadamard form.

We finally remark that the sums in~\eqref{fact1} and~\eqref{fact2} can be understood as
perturbation expansions in the amplitude~$\lambda$ of the harmonic plane wave.
Linearly in~$\lambda$, we recover the usual vacuum polarization as
described by Schwinger in~\cite{schwinger51}.
The higher orders in~$\lambda$ in~\eqref{fact1} and~\eqref{fact2}
can be understood as processes involving multiple photon absorptions and/or emissions.

\section{Outlook} \label{secoutlook}
This is the first paper where the mass oscillation property is proved
for a time-dependent external potential in Minkowski space without decay assumptions at infinity.
Our methods make essential use of the fact that the external field propagates with
the speed of light in a preferred direction. It seems likely that our methods
also apply to pp-space-times (see~\cite[Section~24.5]{stephanietal})
or other space-times involving plane gravitational and electromagnetic waves.

Another possible extension is to include an additional potential~$\Delta \B$ which
does not need to have any symmetries, but which is not too large and
has suitable decay properties at infinity.
In order to treat such a potential, one takes the fermionic signature
operator of Theorem~\ref{thmsig} as the starting point
and treats~$\Delta \B$ as a finite perturbation.
Indeed, since the fermionic signature operator of Theorem~\ref{thmsig} 
has a spectral gap, one could adapt the
methods developed in~\cite{hadamard} for an external potential in Minkowski space.
We expect that the resulting FP state should be again of Hadamard form.

\Thanks {{\em{Acknowledgments:}}
We would like to thank Dirk-Andr{\'e} Deckert for helpful discussions.
We are grateful to the referees for useful suggestions on the manuscript.


\begin{thebibliography}{10}

\bibitem{araki1970quasifree}
H.~Araki, \emph{On quasifree states of {${\rm CAR}$} and {B}ogoliubov
  automorphisms}, Publ. Res. Inst. Math. Sci. \textbf{6} (1970/71), 385--442.

\bibitem{baer+fredenhagen}
C.~B\"ar and K.~Fredenhagen~(eds), \emph{Quantum {F}ield {T}heory on {C}urved
  {S}pacetimes}, Lecture Notes in Physics, vol. 786, Springer Verlag, Berlin,
  2009.

\bibitem{baer+ginoux}
C.~B{\"a}r, N.~Ginoux, and F.~Pf{\"a}ffle, \emph{Wave {E}quations on
  {L}orentzian {M}anifolds and {Q}uantization}, ESI Lectures in Mathematics and
  Physics, European Mathematical Society (EMS), Z\"urich, 2007.

\bibitem{dappiaggiDirac}
C.~Dappiaggi, T.-P. Hack, and N.~Pinamonti, \emph{The extended algebra of
  observables for {D}irac fields and the trace anomaly of their stress-energy
  tensor}, arXiv:0904.0612 [math-ph], Rev. Math. Phys. \textbf{21} (2009),
  no.~10, 1241--1312.

\bibitem{merkl}
D.-A. Deckert, D.~D{\"u}rr, F.~Merkl, and M.~Schottenloher,
  \emph{Time-evolution of the external field problem in quantum
  electrodynamics}, arXiv:0906.0046v2 [math-ph], J. Math. Phys. \textbf{51}
  (2010), no.~12, 122301, 28.

\bibitem{deckert+merkl2}
D.-A. Deckert and F.~Merkl, \emph{External field {QED} on {C}auchy surfaces for
  varying electromagnetic fields}, arXiv:1505.06039 [math-ph], Comm. Math.
  Phys. \textbf{345} (2016), no.~3, 973--1017.

\bibitem{fewster+lang}
C.J. Fewster and B.~Lang, \emph{Pure quasifree states of the {D}irac field from
  the fermionic projector}, arXiv:1408.1645 [math-ph], Class. Quantum Grav.
  \textbf{32} (2015), no.~9, 095001, 30.

\bibitem{fewster2013necessity}
C.J. Fewster and R.~Verch, \emph{The necessity of the {H}adamard condition},
  arXiv:1307.5242 [gr-qc], Class. Quantum Grav. \textbf{30} (2013), no.~23,
  235027, 20.

\bibitem{fierz+scharf}
H.~Fierz and G.~Scharf, \emph{Particle interpretation for external field
  problems in {QED}}, Helv. Phys. Acta \textbf{52} (1979), no.~4, 437--453.

\bibitem{sea}
F.~Finster, \emph{Definition of the {D}irac sea in the presence of external
  fields}, arXiv:hep-th/9705006, Adv. Theor. Math. Phys. \textbf{2} (1998),
  no.~5, 963--985.

\bibitem{light}
\bysame, \emph{Light-cone expansion of the {D}irac sea in the presence of
  chiral and scalar potentials}, arXiv:hep-th/9809019, J. Math. Phys.
  \textbf{41} (2000), no.~10, 6689--6746.

\bibitem{cfs}
\bysame, \emph{The {C}ontinuum {L}imit of {C}ausal {F}ermion {S}ystems},
  arXiv:1605.04742 [math-ph], Fundamental Theories of Physics, vol. 186,
  Springer, 2016.

\bibitem{drum}
F.~Finster and O.~M\"uller, \emph{Lorentzian spectral geometry for globally
  hyperbolic surfaces}, arXiv:1411.3578 [math-ph], Adv. Theor. Math. Phys.
  \textbf{20} (2016), no.~4, 751--820.

\bibitem{hadamard}
F.~Finster, S.~Murro, and C.~R\"oken, \emph{The fermionic projector in a
  time-dependent external potential: Mass oscillation property and {H}adamard
  states}, arXiv:1501.05522 [math-ph], J. Math. Phys. \textbf{57} (2016),
  072303.

\bibitem{rindler}
\bysame, \emph{The fermionic signature operator and quantum states in {R}indler
  space-time}, arXiv:1606.03882 [math-ph] (2016).

\bibitem{finite}
F.~Finster and M.~Reintjes, \emph{A non-perturbative construction of the
  fermionic projector on globally hyperbolic manifolds {I} -- {S}pace-times of
  finite lifetime}, arXiv:1301.5420 [math-ph], Adv. Theor. Math. Phys.
  \textbf{19} (2015), no.~4, 761--803.

\bibitem{infinite}
\bysame, \emph{A non-perturbative construction of the fermionic projector on
  globally hyperbolic manifolds {II} -- {S}pace-times of infinite lifetime},
  arXiv:1312.7209 [math-ph], Adv. Theor. Math. Phys. \textbf{20} (2016), no.~5,
  1007--1048.

\bibitem{fradkin}
E.S. Fradkin, D.M. Gitman, and Sh.M. Shvartsman, \emph{Quantum
  {E}lectrodynamics with {U}nstable {V}acuum}, Springer-Verlag, Berlin, 1991.

\bibitem{hollands+wald}
S.~Hollands and R.M. Wald, \emph{Quantum fields in curved spacetime},
  arXiv:1401.2026 [gr-qc], Phys. Rep. \textbf{574} (2015), 1--35.

\bibitem{hormanderhyp}
L.~H{\"o}rmander, \emph{Lectures on {N}onlinear {H}yperbolic {D}ifferential
  {E}quations}, Math\'ematiques \& Applications (Berlin) [Mathematics \&
  Applications], vol.~26, Springer-Verlag, Berlin, 1997.

\bibitem{john}
F.~John, \emph{Partial {D}ifferential {E}quations}, fourth ed., Applied
  Mathematical Sciences, vol.~1, Springer-Verlag, New York, 1991.

\bibitem{klaus+scharf1}
M.~Klaus and G.~Scharf, \emph{The regular external field problem in quantum
  electrodynamics}, Helv. Phys. Acta \textbf{50} (1977), no.~6, 779--802.

\bibitem{klaus+scharf2}
\bysame, \emph{Vacuum polarization in {F}ock space}, Helv. Phys. Acta
  \textbf{50} (1977), no.~6, 803--814.

\bibitem{rejzner}
K.~Rejzner, \emph{Perturbative {A}lgebraic {Q}uantum {F}ield {T}heory}, Math.
  Phys. Stud., Springer, 2016.

\bibitem{ruijsenaars}
S.N.M. Ruijsenaars, \emph{Charged particles in external fields. {I}.
  {C}lassical theory}, J. Mathematical Phys. \textbf{18} (1977), no.~4,
  720--737.

\bibitem{sahlmann2001microlocal}
H.~Sahlmann and R.~Verch, \emph{Microlocal spectrum condition and {H}adamard
  form for vector-valued quantum fields in curved spacetime},
  arXiv:math-ph/0008029, Rev. Math. Phys. \textbf{13} (2001), no.~10,
  1203--1246.

\bibitem{schwinger51}
J.~Schwinger, \emph{On gauge invariance and vacuum polarization}, Physical Rev.
  (2) \textbf{82} (1951), 664--679.

\bibitem{shale+stinespring}
D.~Shale and W.F. Stinespring, \emph{Spinor representations of infinite
  orthogonal groups}, J. Math. Mech. \textbf{14} (1965), 315--322.

\bibitem{stephanietal}
H.~Stephani, D.~Kramer, M.~MacCallum, C.~Hoenselaers, and E.~Herlt, \emph{Exact
  solutions of {E}instein's field equations}, second ed., Cambridge Monographs
  on Mathematical Physics, Cambridge University Press, Cambridge, 2003.

\bibitem{treude}
J.-H. Treude, \emph{Estimates of {M}assive {D}irac {W}ave {F}unctions near
  {N}ull {I}nfinity}, Dissertation, Universit\"at Regensburg,
  http://epub.uni-regensburg.de/32344/, 2015.

\bibitem{volkow}
D.M. Volkow, \emph{{\"Uber eine Klasse von L\"osungen der Diracschen
  Gleichung}}, Z. Physik \textbf{94} (1935), 250--260.

\bibitem{waldQFT}
R.M. Wald, \emph{Quantum {F}ield {T}heory in {C}urved {S}pacetime and {B}lack
  {H}ole {T}hermodynamics}, Chicago Lectures in Physics, University of Chicago
  Press, Chicago, IL, 1994.

\end{thebibliography}
\providecommand{\bysame}{\leavevmode\hbox to3em{\hrulefill}\thinspace}
\providecommand{\MR}{\relax\ifhmode\unskip\space\fi MR }
\providecommand{\MRhref}[2]{%
  \href{http://www.ams.org/mathscinet-getitem?mr=#1}{#2}
}
\providecommand{\href}[2]{#2}

\end{document}